\documentclass[aps,twocolumn,superscriptrddress,10pt]{revtex4-1}
\pdfoutput=1

\usepackage{color}
\usepackage[latin1]{inputenc}
\usepackage{graphicx}
\usepackage{amsmath,amssymb}
\usepackage{booktabs}
\usepackage{hyperref}
\usepackage{braket}
\hypersetup{colorlinks=true,
						linkcolor=red,
						anchorcolor=blue,
						citecolor=blue,
						filecolor=blue,
						menucolor=blue,
						urlcolor=blue
}

\renewcommand{\vec}[1]{{\boldsymbol{#1}}}
\renewcommand\Re{\operatorname{Re}}
\renewcommand\Im{\operatorname{Im}}

\begin{document}
	
\title{Experimental realisation of the topological Haldane model with ultracold fermions}

\date{\today}

\author{
Gregor~Jotzu, Michael~Messer, R\'emi~Desbuquois, Martin~Lebrat, Thomas~Uehlinger, Daniel~Greif \& Tilman~Esslinger }

\affiliation{Institute for Quantum Electronics, ETH Zurich, 8093 Zurich, Switzerland}

\date{\today}

\maketitle

{\bf 
The Haldane model on the honeycomb lattice is a paradigmatic example of a Hamiltonian featuring topologically distinct phases of matter \cite{Haldane1988}. It describes a mechanism through which a quantum Hall effect can appear as an intrinsic property of a band-structure, rather than being caused by an external magnetic field \cite{Chang2013}. Although an implementation in a material was considered unlikely, it has provided the conceptual basis for theoretical and experimental research exploring topological insulators and superconductors \cite{Kane2005,Konig2007,Hsieh2008,Chang2013,Hasan2010}. Here we report on the experimental realisation of the Haldane model and the characterisation of its topological band-structure, using ultracold fermionic atoms in a periodically modulated optical honeycomb lattice. The model is based on breaking time-reversal symmetry as well as inversion symmetry. The former is achieved through the introduction of complex next-nearest-neighbour tunnelling terms, which we induce through circular modulation of the lattice position \cite{Oka2009}. For the latter, we create an energy offset between neighbouring sites \cite{Tarruell2012}. Breaking either of these symmetries opens a gap in the band-structure, which is probed using momentum-resolved interband transitions. We explore the resulting Berry-curvatures of the lowest band by applying a constant force to the atoms and find orthogonal drifts analogous to a Hall current. The competition between both broken symmetries gives rise to a transition between topologically distinct regimes. By identifying the vanishing gap at a single Dirac point, we map out this transition line experimentally and quantitatively compare it to calculations using Floquet theory without free parameters. We verify that our approach, which allows for dynamically tuning topological properties, is suitable even for interacting fermionic systems. Furthermore, we propose a direct extension to realise spin-dependent topological Hamiltonians.
}

\begin{figure}
\includegraphics[width=1\columnwidth]{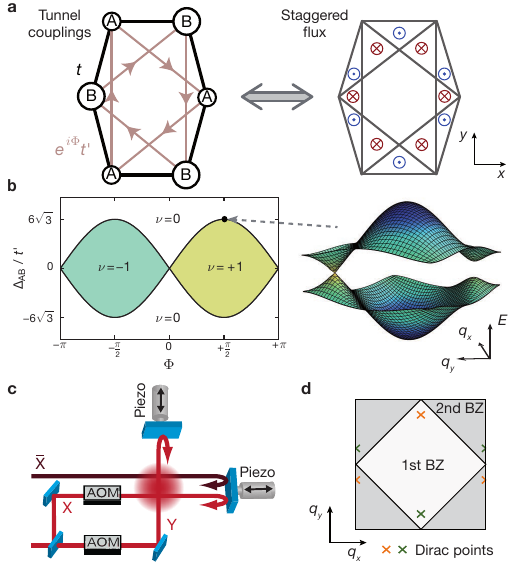}
\caption{
{\bf The Haldane model.} 
{\bf a,} 
Tight-binding model of the honeycomb lattice realised in the experiment. An energy offset $\Delta_{\mathrm{AB}}$ between sublattice A and B breaks inversion symmetry (IS). Nearest-neighbour tunnel couplings $t_{ij}$ are real-valued, whereas next-nearest-neighbour tunnelling $e^{i\Phi_{ij}} t'_{ij}$ carries tunable phases indicated by arrows. $i$ and $j$ indicate the indices of the connected lattice sites. For a detailed discussion of anisotropies and higher order tunnelling terms see Supplementary Material. Corresponding staggered magnetic fluxes (sketched on the right) sum up to zero but break time-reversal symmetry (TRS). 
{\bf b,} 
Topological regimes of the Haldane model, for isotropic $t$, $t'$ and $\Phi$. The trivial (Chern number $\nu = 0$) and Chern-insulating ($\nu = \pm1$) regimes are connected by topological transitions (black lines), where the band-structure (shown on the right) becomes gapless at a single Dirac point. 
{\bf c,} 
Laser beam set-up forming the optical lattice. $\overline{\mathrm{X}}$ is frequency-detuned from the other beams. Piezo-electric actuators sinusoidally modulate the retro-reflecting mirrors, with a controllable phase difference $\varphi$. Acousto-optic modulators (AOMs) ensure the stability of the lattice geometry  (Methods). 
{\bf d,} 
Resulting Brillouin-zones featuring two Dirac-points.
}
\label{fig:haldane-scheme}
\end{figure}

In a honeycomb lattice symmetric under time-reversal and inversion, the two lowest bands are connected at two Dirac points. Each broken symmetry leads to a gapped energy-spectrum. F. D. M. Haldane realised that the resulting phases are topologically distinct: A broken inversion symmetry (IS), caused by an energy offset between the two sublattices, leads to a trivial band-insulator at half-filling. Time-reversal symmetry (TRS) can be broken by complex next-nearest-neighbour tunnel couplings (Fig. \ref{fig:haldane-scheme}a). The corresponding staggered magnetic fluxes sum up to zero in one unit-cell, thereby preserving the translation symmetry of the lattice. This gives rise to a topological Chern-insulator, where a non-zero Hall conductance appears despite the absence of a net magnetic field \cite{Haldane1988,Chang2013}. When both symmetries are broken, a topological phase transition connects two regimes with a distinct topological invariant, the Chern number, which changes from 0 to $\pm 1$, see Fig. \ref{fig:haldane-scheme}b. There, the gap closes at a single Dirac point. These transitions have attracted great interest, as they cannot be described by Landau's theory of phase transitions, owing to the absence of a changing local order parameter \cite{Hasan2010}.

A crucial experimental challenge for the realisation of the Haldane model is the creation of complex next-nearest-neighbour tunnelling. Here we show that this becomes possible with ultracold atoms in optical lattices periodically modulated in time. So far, pioneering experiments with bosons showed a renormalization of existing tunnelling amplitudes in one dimension \cite{Dunlap1986,Lignier2007}, and were extended to control tunnelling phases \cite{Struck2012,Jimenez2012} and higher-order tunnelling \cite{Parker2013}. In higher dimensions this allowed for studying phase transitions \cite{Zenesini2009,Struck2011}, and topologically trivial staggered fluxes were realised \cite{Aidelsburger2011,Struck2013}. Furthermore, uniform flux configurations were observed using rotation and laser-assisted tunnelling \cite{Williams2010, Aidelsburger2013}, although for the latter method, heating seemed to prevent the observation of a flux in some experiments \cite{Miyake2013}. In a honeycomb lattice, a rotating force can induce the required complex tunnelling, as recognised by T. Oka and H. Aoki \cite{Oka2009}. Using arrays of coupled waveguides, a classical version of this proposal was used to study topologically protected edge modes in the inversion-symmetric regime \cite{Rechtsman2013}. We access the full parameter space of the Haldane model using a fermionic quantum gas, by extending the proposal to elliptical modulation of the lattice position and additionally breaking IS through a deformation of the lattice geometry.

The starting point of our experiment is a non-interacting, ultracold gas of $4\times 10^4$ to $6\times 10^4$ fermionic $^{40}$K atoms prepared in the lowest band of a honeycomb optical lattice created by several laser beams at wavelength $\lambda = 1064 $\,nm, arranged in the $x-y$ plane as depicted in Fig. \ref{fig:haldane-scheme}c and detailed in \cite{Tarruell2012}. The two lowest bands have a total bandwidth of $h \times 3.9(1)$\,kHz, with a gap of $h \times 5.4(2)$\,kHz to the next higher band, and contain two Dirac points at opposite quasi-momenta, see Fig. \ref{fig:haldane-scheme}d. Here $h$ denotes Planck's constant. After loading the atoms into the honeycomb lattice, we ramp on a sinusoidal modulation of the lattice position along the $x$ and $y$ directions with amplitude $0.087(1)\lambda$, frequency $4.0$\,kHz and phase difference $\varphi$. This gives access to linear ($\varphi=0^{\circ}$ or $180^{\circ}$), circular ($\varphi=\pm 90^{\circ}$) and elliptical trajectories.

\begin{figure*}
\includegraphics[width=2\columnwidth]{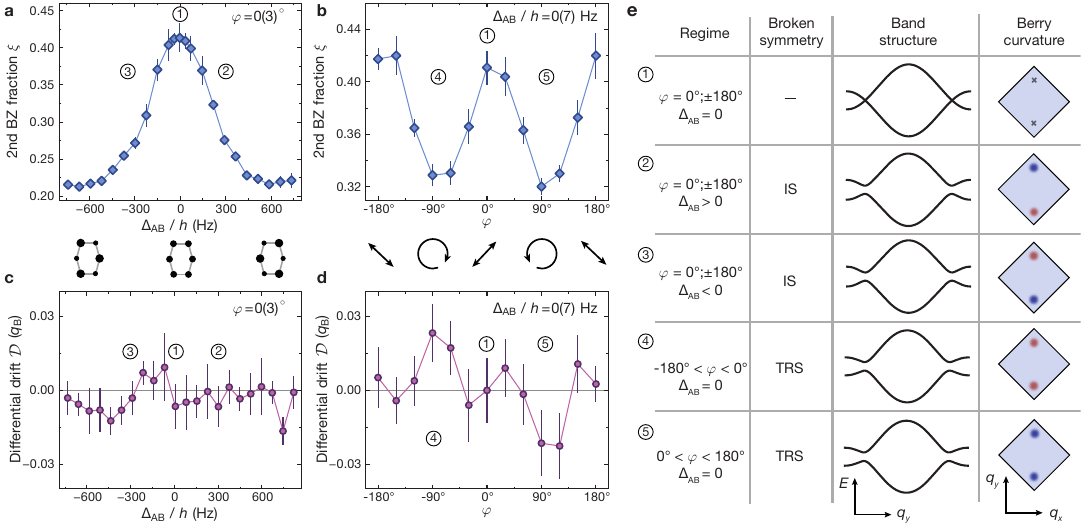}
\caption{
{\bf Probing gaps and Berry-curvature.} 
{\bf a+b,} 
Fraction of atoms in the second band $\xi$ after one Bloch-oscillation in the $q_x$-direction. We break either IS ({\bf a}) by introducing a sublattice offset $\Delta_{\mathrm{AB}}$ or TRS ({\bf b}) \textit{via} elliptical modulation (see diagrams below). This corresponds to scanning either of the two axes of the Haldane model. A gap opens at both Dirac points, given by $\vert\Delta_{\mathrm{AB}}\vert$ or $\vert\Delta_{\mathrm{T}}^{\mathrm{max}} \sin\left(\varphi\right)\vert$, respectively, thereby reducing $\xi$. 
{\bf c+d,} 
Differential drift $\cal{D}$ obtained from Bloch-oscillations in opposite $q_y$-directions. For broken IS ({\bf c}), opposite Berry-curvatures at the two Dirac points cancel each other, whilst for broken TRS ({\bf d}) the system enters the topological regime, where opposite drifts for $\varphi \gtrless 0$ are expected. Data show mean $\pm$ s.d. of at least 6({\bf a-c}) or 21({\bf d}) measurements and $q_{\mathrm B}=2\pi/\lambda$.
{\bf e,} 
Sketches illustrating gaps and Berry-curvature in different regimes. Red (blue) indicates positive (negative) Berry-curvature. 
}
\label{fig:observables}
\end{figure*}

The effective Hamiltonian of our system in the phase-modulated honeycomb lattice is computed using analytical and numerical Floquet theory (See Methods and Supplementary Material for a detailed discussion). It is well described by the Haldane model \cite{Haldane1988}
\begin{equation}
\hat{H} = 
\sum_{\langle ij\rangle}t_{ij} \hat{c}_i^{\dagger}\hat{c}_j + 
\sum_{\langle\langle ij\rangle\rangle} e^{i\Phi_{ij}}t'_{ij} \hat{c}_i^{\dagger}\hat{c}_j + 
\Delta_{\mathrm{AB}} \sum_{i \in A} \hat{c}_i^{\dagger}\hat{c}_i,
\label{eq:HaldaneHam}
\end{equation}
where $t_{ij}$ and $t'_{ij}$ are real-valued nearest- and next-nearest-neighbour tunnelling amplitudes, and the latter contain additional complex phases $\Phi_{ij}$ defined along the arrows shown in Fig. \ref{fig:haldane-scheme}a. The fermionic creation and annihilation operators are denoted by $\hat{c}^{\dagger}_i$ and $\hat{c}_i$. The energy offset $\Delta_{\mathrm{AB}}\gtrless 0$ between sites of the $A$ and $B$ sublattice breaks IS and opens a gap $\vert\Delta_{\mathrm{AB}}\vert$ \cite{Tarruell2012}. TRS can be broken by changing $\varphi$. This controls the imaginary part of the next-nearest-neighbour tunnelling, whereas its real part as well as $ t_{ij}$ and $\Delta_{\mathrm{AB}}$ are mostly unaffected (Methods). Breaking only TRS opens an energy gap $\vert\Delta_{\mathrm{T}}\vert$ at the Dirac points given by a sum of the imaginary part of the three next-nearest-neighbour tunnel couplings connecting the same sublattice 
\begin{equation}
\Delta_{\mathrm{T}} = 
- \sum_{l} w_{l}t'_{l}\sin\left(\Phi_{l}\right) = 
\Delta_{\mathrm{T}}^{\mathrm{max}} \sin(\varphi)  ,
\label{eq:tr-gap}
\end{equation}
with weights $w_l$ of order unity, which depend on the position of the Dirac points in the Brillouin zone. The sum is taken over the different types of next-nearest-neighbour bond, and the origin of the second equality is discussed in the Supplementary Information. Circular modulation ($\varphi=\pm90^{\circ}$) leads to a maximum gap ($h\times 88^{+10}_{-34}$\,Hz for our parameters), whereas the gap vanishes for linear modulation ($\varphi=0^{\circ},\pm180^{\circ}$), where TRS is preserved.

We will first present measurements which confirm that breaking either symmetry is sufficient to open a gap in the band structure. For this, we restrict ourselves to either $\varphi=0^\circ$ or $\Delta_{\mathrm{AB}}=0$, corresponding to the two axes of the Haldane diagram of Fig. \ref{fig:haldane-scheme}b. 
Subsequently we will present measurements in which we explore the topology of the lowest band in the same parameter regime by probing the Berry-curvature.
In order to probe the opening of gaps in the system, we drive Landau-Zener transitions through the Dirac points \cite{Tarruell2012, Lim2012}. Applying a constant force along the $x$-direction by means of a magnetic field gradient causes an energy offset $E/h = 103.6(1)$\,Hz per site, thereby inducing a Bloch oscillation. After one full Bloch cycle the gradient is removed and the fraction of atoms $\xi$ in the second band is determined using a band-mapping procedure (Methods). For broken IS, a gap given by $\vert \Delta_\mathrm{AB} \vert$ opens at both Dirac points. In this case, $\xi$ reaches a maximum at $ \Delta_\mathrm{AB} = 0$ which indicates a vanishing energy gap, and decays symmetrically around this point as expected, see Fig. \ref{fig:observables}a. In the case of broken TRS (Fig. \ref{fig:observables}b), a reduction in transfer versus modulation phase is observed. This signals an opening gap, which is found to be largest for circular modulation, as expected from Eq. (\ref{eq:tr-gap}).

Breaking either IS or TRS gives rise to similar, gapped band structures which remain point-symmetric around quasi-momentum $\textbf{q}=0$. However, the energy spectrum itself is not sufficient to reveal the different topology of the band, which is given by the associated eigenstates. These are characterized by a local geometrical property: the Berry-curvature $\Omega(\textbf{q})$ \cite{Hasan2010}. In $\textbf{q}$-space, $\Omega(\textbf{q})$ is analogous to a magnetic field and corresponds to the geometric phase picked up along an infinitesimal loop. When only IS is broken, the Berry-curvature is point-antisymmetric, and its sign inverts for opposite $\Delta_{\mathrm{AB}}$, see Fig. \ref{fig:observables}e. The spread of $\Omega(\textbf{q})$ increases with the size of the gap. Its integral over the first Brillouin-zone , the Chern number $\nu$, is zero, corresponding to a topologically trivial system. However, with only TRS broken, $\nu=\pm1$, $\Omega(\textbf{q})$ is point-symmetric, and its sign changes when reverting the rotation direction of the lattice modulation. 

In order to determine the topology of the lowest band, we move the atoms along the y-direction such that their trajectories sample the regions where the Berry-curvature is concentrated, and record their final position. As atoms move through regions of $\textbf{q}$-space with non-zero curvature, they acquire an orthogonal velocity proportional to the applied force and $\Omega(\textbf{q})$ \cite{Chang1995,Dudarev2004,Price2012,Dauphin2013}. The underlying harmonic confinement caused by the laser beams in the experiment couples real and momentum-space, meaning that a displacement in real space leads to a drift in quasi-momentum. We apply a gradient of $\Delta E/h = 114.6(1)$\,Hz per site and measure the center of mass of the quasi-momentum distribution in the lowest band after one full Bloch cycle. As the velocity caused by the Berry-curvature inverts when inverting the force, we subtract the result for the opposite gradient to obtain the differential drift $\cal{D}$. This quantity is more suitable for distinguishing trivial from non-trivial Berry-curvature distributions than the response to a single gradient (Methods) \cite{Price2012}.
The latter does however provide information about the local Berry-curvature and is shown in Extended Data Figure 2.
When breaking only IS, we observe that $\cal{D}$ vanishes and is independent of $\Delta_\mathrm{AB}$, as the Berry-curvature is point-antisymmetric, see Fig. \ref{fig:observables}c. 
In contrast, when only TRS is broken, we explore the topological regime of the Haldane model with $\Delta_{\mathrm{AB}}=0$.  A differential drift is observed for $\varphi=90^{\circ}$, which, as expected, is opposite for $\varphi=-90^{\circ}$, see Fig. \ref{fig:observables}d and \ref{fig:driftphasediagram}c. This is a direct consequence of the Berry-curvature being point-symmetric, with its sign given by the rotation direction of the lattice modulation. In fact, here a non-zero $\cal{D}$ can only originate from a non-zero integrated Berry-curvature (Methods). 
As the modulation becomes linear, the drift disappears. This is smoothed by the increased transfer to the higher band when the gap becomes smaller, which predominantly affects atoms that would experience the strongest Berry-curvature.
These observations are qualitatively confirmed by semiclassical simulations shown in Extended Data Figure 1.

Within the Haldane model, the competition of simultaneously broken TRS and IS is of particular interest, as it features a topological transition between a trivial band insulator and a Chern-insulator. In this regime, both the band-structure and Berry-curvature are no longer point-symmetric and the energy gap $G_{\pm}$ at the two Dirac points is given by
\begin{equation}
G_{\pm} = \vert \Delta_\mathrm{AB} \pm \Delta_{\mathrm{T}}^{\mathrm{max}}\cdot \sin(\varphi) \vert.
\label{eq:gap}
\end{equation}
On the transition lines the system is gapless with one closed and one gapped Dirac point, $G_{+} = 0$ or $G_{-} = 0$. We will now discuss measurements in which we extend the parameter regime to allow for the simultaneous breaking of both symmetries.

\begin{figure}
\includegraphics[width=1\columnwidth]{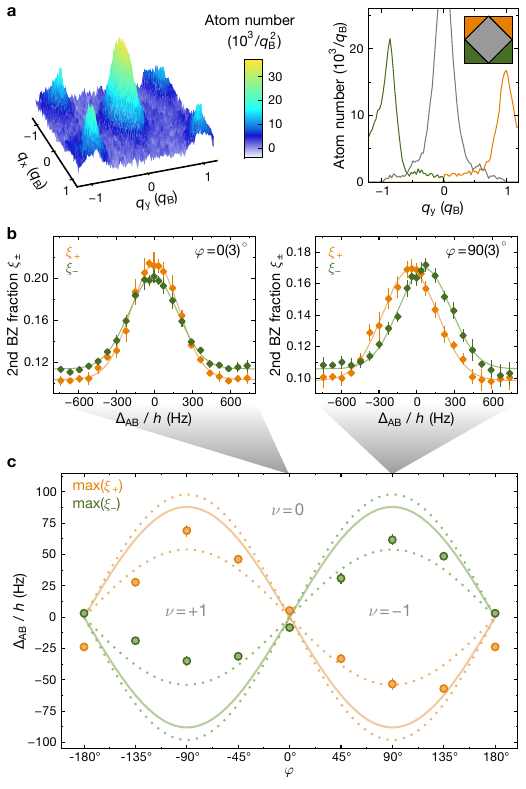}
\caption{%
{\bf Mapping out the transition line.} 
{\bf a,} 
Atomic quasi-momentum distribution (averaged over 6 runs) after one Bloch-oscillation for $\varphi = +90^{\circ}, \Delta_{\mathrm{AB}}/h = 292(7)$\,Hz. A line-sum along $q_x$ shows the atomic density in the first Brillouin-zone in grey; atoms transferred at the upper (lower) Dirac point are shown in orange (green) throughout. The fraction of atoms in the second Brillouin-zone differs for $q_y \gtrless 0$. This is a direct consequence of simultaneously broken IS and TRS, which allows for band-structures that are not point-symmetric. {\bf b,} Fractions of atoms $\xi_{\pm}$ in each half of the second Brillouin-zone. For linear modulation (left) the gap vanishes at $\Delta_{\mathrm{AB}} = 0$ for both Dirac points, whilst for circular modulation (right) it vanishes at opposite values of $\Delta_{\mathrm{AB}}$. Gaussian fits (solid lines) are used to find the maximum transfer, which signals the topological transition. Data are mean $\pm$ s.d. of at least 6 measurements. {\bf c,} Solid lines show the theoretically computed topological transitions, without free parameters. Dotted lines represent the uncertainty of the maximum gap $\vert\Delta_{\mathrm{T}}^{\mathrm{max}} \vert/h= 88^{+10}_{-34}\mathrm{Hz}$, originating from the uncertainty of the lattice parameters. Data are the points of maximum transfer for each Dirac point, $\pm$ fit error, obtained from measurements as in {\bf b} for various $\varphi$. Data points for $\varphi=\pm 180^{\circ}$ correspond to the same measurements.
Between the lines, the system is in the topologically non-trivial regime.
}
\label{fig:imbalance}
\end{figure}

\begin{figure}
\includegraphics[width=1\columnwidth]{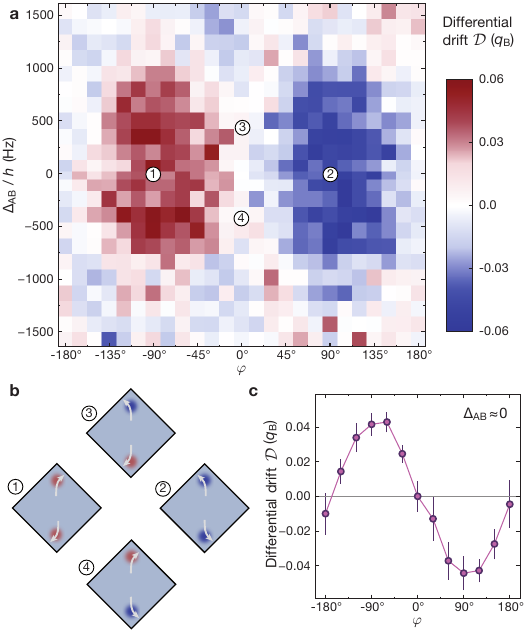}
\caption{%
{\bf Drift measurements.} 
{\bf a,} 
Differential drift $\cal{D}$ in quasi-momentum. Each pixel corresponds to at least one pair of measurements, where the modulation frequency was set to $3.75$\,kHz. Data points for $\varphi=\pm 120^\circ$, $\Delta_{\mathrm{AB}}/h=503(7)$\,Hz were not recorded and are interpolated. 
{\bf b,}
All topological regimes are explored and the expected momentum-space drifts caused by the Berry-curvature are sketched for selected parameters. 
{\bf c,} 
Cut along the $\Delta_{\mathrm{AB}}/h=15(7)$\,Hz line. Data show mean $\pm$ s.d. of at least 6 pairs of measurements.
}
\label{fig:driftphasediagram}
\end{figure}

We map out the transition by measuring the transfer $\xi_{\pm}$ for each Dirac point separately, see Fig. \ref{fig:imbalance}a. $\xi_+$ ($\xi_-$) is the fraction of atoms occupying the upper (lower) half of the second Brillouin-zone after one Bloch oscillation along the $x$-direction. We observe a difference between $\xi_{+}$ and $\xi_{-}$ which shows that the band structure is no longer point-symmetric, allowing for the parity anomaly predicted by F. D. M. Haldane \cite{Haldane1988}. When the topology of the band changes, the gap at one of the Dirac point closes. We identify the closing of a gap with the point of maximum measured transfer when varying $\Delta_\mathrm{AB}$. For $\varphi=0^{\circ}$ we find, as expected for preserved TRS, that the maxima of both $\xi_+$ and $\xi_-$ coincide, see Fig. \ref{fig:imbalance}b. The maxima are shifted in opposite directions for $\varphi=90^{\circ}$, showing that the minimum gap for each Dirac point occurs at different values of $\Delta_\mathrm{AB}$. In between these values the system is in the topologically non-trivial regime. We explore the position of each maximum for varying $\varphi$ and find opposite shifts for negative $\varphi$ as predicted by Eq. (\ref{eq:gap}) using no free parameters, see Fig. \ref{fig:imbalance}c.

In Figure 4 we show the measured differential drift $\cal{D}$ for all topological regimes, allowing for simultaneously broken IS and TRS. Here, we reduce the modulation frequency to $3.75$\,kHz where the signal-to-noise ratio of $\cal{D}$ is larger, but which is less suited for a quantitative comparison of the transfer $\xi$, as the lattice modulation ramps are expected to be less adiabatic. $\cal{D}$ is non-zero only for broken TRS and shows the expected antisymmetry with $\varphi$ and symmetry with $\Delta_{\mathrm{AB}}$. For large $\Delta_{\mathrm{AB}}$, deep inside the topologically trivial regime, $\cal{D}$ vanishes for all $\varphi$. For smaller $\Delta_{\mathrm{AB}}$, the differential drift shows precursors of the regimes with non-zero Chern number: non-zero values of $\cal{D}$ extend well beyond the transition lines when IS and TRS are both broken. Semiclassical simulations (see Extended Data Figure 1c) suggest that the main contribution to this effect arises from the transfer to the higher band.

Extending our work to interacting systems requires sufficiently low heating. We investigate this with a repulsively interacting spin-mixture in the honeycomb lattice previously used for studying the fermionic Mott insulator \cite{Uehlinger2013}. We measure the entropy in the Mott insulating regime by loading into the lattice and reverting the loading procedure (see Methods and Extended Data Figure 3). The entropy increase is only $25\%$ larger than without modulation. This opens the possibility of studying topological models with interactions \cite{Varney2010} in a controlled and tunable way. For example, lattice modulation could be used to create topological flat bands, which have been suggested to give rise to interaction-induced fractional Chern-insulators \cite{Neupert2011, Grushin2014}. Furthermore, our approach of periodically modulating the system can be directly extended to engineer Hamiltonians with spin-dependent tunnelling amplitudes and phases (Methods). This can be achieved by modulating a magnetic field gradient, which leads to spin-dependent oscillating forces owing to the differential Zeeman shift. For example, time-reversal symmetric topological Hamiltonians, such as the Kane-Mele model \cite{Kane2005}, can be implemented by simultaneously modulating the lattice on one axis and a magnetic field gradient on the other.

\setlength{\parindent}{0pt}

\textbf{Acknowledgements} 
We would like to thank Hideo Aoki for pointing out their proposal in the context of optical lattices and Nigel Cooper, Sebastian Huber, Leticia Tarruell, Lei Wang and Alessandro Zenesini for insightful discussions. We acknowledge SNF, NCCR-QSIT and SQMS (ERC advanced grant) for funding.

\textbf{Author Contributions} 
The data were measured by G.J., M.M., R.D. and D.G. and analysed by G.J., M.M., R.D., T.U. and D.G. The theoretical framework was developed by G.J. and M.L. All work was supervised by T.E. All authors contributed to planning the experiment, discussions and the preparation of the manuscript.

\textbf{Author Information} 
The authors declare no competing financial interests. Correspondence and requests for materials should be addressed to T.E. (esslinger@phys.ethz.ch).

\section*{Methods}

%
    %

\setcounter{section}{0}
\setcounter{subsection}{0}
\setcounter{figure}{0}
\setcounter{equation}{0}
\setcounter{table}{0}
\setcounter{NAT@ctr}{0}

\renewcommand{\thetable}{M\arabic{table}}
\renewcommand{\theequation}{M\arabic{equation}}

\renewcommand{\figurename}[1]{EXTENDED DATA FIG. }
\renewcommand{\thefigure}{ED\arabic{figure}}

\subsection{Spin polarised Fermi gas}

After sympathetic cooling with $^{87}$Rb in a magnetic trap, $1\times10^6$
fermionic $^{40}$K atoms are transferred into an optical dipole trap operating
at a wavelength of $826\,\text{nm}$. A balanced spin mixture of the
$|F,m_F\rangle=|9/2,-9/2\rangle$ and $|9/2,-7/2\rangle$ Zeeman states, where $F$ denotes the hyperfine manifold and $m_F$ the magnetic sub-level, is
evaporatively cooled at a magnetic field of $197.6(1)$\,G, in the vicinity of
the Feshbach resonance located at $202.1$\,G. We obtain typical temperatures of
$0.2\,T_F$, where $T_F$ is the Fermi temperature. The
field is subsequently reduced and a magnetic field gradient is used to
selectively remove the $|9/2,-7/2\rangle$ component, while levitating the atoms
in the $|9/2,-9/2\rangle$ state against gravity. 

\subsection{Honeycomb optical lattice}

This polarised Fermi gas is loaded into the optical lattice
within $200$\,ms using an $S$-shaped intensity ramp, and the dipole trap is
subsequently turned off in $100$\,ms. The lattice potential is given by
\cite{Tarruell2012}
\begin{eqnarray}
V(x,y,z)&=&-V_{\overline{\mathrm{X}}}\cos^2(k_{\mathrm{L}} x+\theta/2)-V_{\mathrm{X}} \cos^2(k_{\mathrm{L}} x)\\
&&-V_{\mathrm{Y}} \cos^2(k_{\mathrm{L}} y) -V_{\tilde{\mathrm{Z}}} \cos^2(k_{\mathrm{L}} z)\nonumber 
\\
&&-2\alpha \sqrt{V_{\mathrm{X}}V_{\mathrm{Y}}}\cos(k_{\mathrm{L}} x)\cos(k_{\mathrm{L}} y)\cos\varphi_{\mathrm{L}}, \nonumber
\label{eq:lattice}
\end{eqnarray}
where $V_{\overline{\mathrm{X}},\mathrm{X},\mathrm{Y},\tilde{\mathrm{Z}}}$ are the single-beam lattice depths and
$k_{\mathrm{L}}=2\pi/\lambda$. Gravity points along the negative
$y$-direction. To control the energy offset $\Delta_{\mathrm{AB}}$, we vary
$\theta$ around $\pi$ by changing the frequency detuning $\delta$ between the $\overline{\mathrm{X}}$ 
and the $\mathrm{X}$ (which has the same frequency as $\mathrm{Y}$) beams using an acousto-optic modulator, see Fig. \ref{fig:haldane-scheme}c. 
The point where $\Delta_{\mathrm AB}=0$ is determined from transfer measurements in a static lattice with an accuracy of 7 Hz.
The phase $\varphi_{\mathrm{L}}$ is
stabilised to $0.0(3)^\circ$ using a heterodyne interferometer \cite{Tarruell2012}, and the
visibility of the interference pattern is $\alpha=0.81(1)$. We minimise the intensity imbalance between the incoming and reflected lattice beams in the $x-y$
plane such that the remaining imbalance between left and right vertical
tunnelling is less than $0.3\%$, as determined from Raman-Nath diffraction on a $^{87}$Rb Bose-Einstein condensate. The final lattice depths are set to
$V_{\overline{\mathrm{X}},\mathrm{X},\mathrm{Y},\tilde{\mathrm{Z}}}=[5.0(3),0.45(2),2.3(1),0]\,E_{\mathrm{R}}$, where $E_{\mathrm{R}}=h^2 /2 m\lambda^2$, and $m$ denotes the mass of $^{40}$K. Using a Wannier
function calculation\cite{Uehlinger2013}, we extract the corresponding tight-binding 
parameters $t_{0,1,2,3}/h=[-746(81),-527(17),-527(17),-126(7)]$\,Hz, for the
horizontal, the left and right vertical nearest-neighbour tunnelling links, and
the horizontal link across the honeycomb cell, respectively (see Supplementary
Material). This results in a bandwidth of $W/h=3.9(1)$\,kHz. The amplitudes
for the next-nearest-neighbour tunnel couplings are $t'_{1,2,3}/h=[14,14,61]$\,Hz and
do not affect the gaps and the topological transition line. The nearest-neighbour tunnelling
is renormalized in the modulated lattice, which decreases the effective bandwidth to
$W_{\mathrm{eff}}/h=3.3(1)$\,kHz. All experiments are carried out in the
presence of a weak underlying harmonic confinement with trapping frequencies
$\omega_{x,y,z}/2\pi=[14.4(6),30.2(1),29.3(3)]$\,Hz, which originates from the Gaussian intensity profiles of the red-detuned lattice beams. The lattice depths are
calibrated using Raman-Nath diffraction on a $^{87}$Rb Bose-Einstein condensate. 
To determine $\alpha$, we drive
quasi-momentum resolved interband transitions for a spin polarised Fermi gas
loaded into a chequerboard lattice by periodically modulating the lattice depth and measure the band gap at ${\bf q}=0$.

\subsection{Modulation of the optical lattice}

The two mirrors used for retro-reflecting the lattice beams are mounted on
piezo-electric actuators, which allow for a controlled phase shift of the
reflected beams with respect to the incoming lattice beams. To fix the geometry of the
lattice, the relative phase $\varphi_{\mathrm{L}}$ of the two orthogonal retro-reflected
beams $\mathrm{X}$ and $\mathrm{Y}$ is actively stabilised to $\varphi_{\mathrm{L}} = 0^\circ$. In order to maintain this
phase relation during modulation, the phase of the respective incoming
beams is modulated at the same frequency as the piezo-electric
actuators using acousto-optical modulators. In addition, this phase modulation provides a direct calibration of
the amplitude and relative phase of the mirror displacement. The calibration is
confirmed by measuring both the reduction of tunnelling \cite{Lignier2007} and
the effective atomic mass around $\textbf{q}=0$ in a modulated simple cubic lattice.

The modulation is turned on as follows: the atoms are loaded into a lattice
with $30\%$ larger single-beam lattice depths than the final values used for the
actual measurements. This suppresses resonant transfer of atoms to
higher bands. The modulation amplitude is then linearly increased within
$20$\,ms to reach a normalised drive of $K_0=0.7778$, where
$K_0=\pi^2\,(A/\lambda)\,(\hbar\,\omega/E_{\mathrm{R}})$ with $A$ the amplitude of the motion, $\omega/2\pi$ the modulation frequency and $\hbar=h/2\pi$. The time-dependence of the lattice position $\textbf{r}_{\text{lat}}(t)$ is then given by 
\begin{equation}
\textbf{r}_{\text{lat}} = -A \Big( \cos(\omega t) \textbf{e}_x + \cos(\omega t - \varphi) \textbf{e}_y \Big),
\end{equation}
where $\textbf{e}_x$ and $\textbf{e}_y$ denote the real-space unit vectors along the 
$x$- and $y$-direction. 
The phase $\varphi$ is set with an accuracy of $3^\circ$.
When using different modulation frequencies,
we keep $K_0$ constant. The
lattice depths are then finally reduced in $10$\,ms to their final values. We have independently confirmed that the modulation frequency of
$\omega/2\pi=4$\,kHz exceeds the combined static bandwidth of the two lowest bands using lattice phase-modulation spectroscopy.  

We have verified that our experimental findings are not affected by the global phase of the lattice modulation or changes in the total modulation time smaller than a modulation period. This means that a time-independent effective Hamiltonian can safely be used. Additionally, we have varied the time-scale on which the modulation is ramped on and off to confirm that these ramps are sufficiently slow and the measurements are not affected by switch-on or switch-off effects.

\subsection{Detection}

After one Bloch cycle lasting $9.85$\,ms ($8.72$\,ms) for $x$ ($y$) oscillations, the
lattice modulation amplitude is linearly lowered to zero within $2$\,ms. The
quasi-momentum distribution of the atoms is then probed using a band mapping
technique, where all lattice beams are ramped down within $500\,\mu$s, which is much shorter than the time-scales of the harmonic trap, meaning that the original $\textbf{q}$-space distribution is conserved~\cite{Tarruell2012}. An absorption image of the atomic distribution is then recorded after $15$\,ms of ballistic expansion. The fraction of atoms per band is determined
by integrating the atomic density in the corresponding Brillouin-zone in the
absorption image. The size of the Brillouin zone is independently calibrated by using Bloch oscillations of a non-interacting Fermi gas in a one-dimensional lattice. 
Owing to the residual non-adiabaticity of the lattice ramps, $16\%$ 
($14\%$) of the atoms are detected in the second band and $21\%$ ($8\%$) in even 
higher bands after loading the lattice and before the Bloch cycle including (not including) the linear ramps of the modulation. 

For the drift measurements, the displacement of the atoms with
respect to the position before the Bloch oscillation is obtained by calculating
the center of mass within the first Brillouin-zone. The differential drift is
then calculated as the difference of the recorded drift for oscillations along
the positive and negative $q_y$-direction. We already observe
individual drifts when only breaking IS, which constitutes a measurement of a local Berry-curvature (see below and Extended Data Figure \ref{fig:edf2}). The drift depends strongly on the size of the energy gap, which supports our explanation for the precursors of the topological phase in Fig. \ref{fig:driftphasediagram}.

\subsection{Effective band-structure calculations}

The effective Hamiltonian $\hat{H}_\mathrm{eff}$ is given by a logarithm of the time-evolution operator for one modulation period.
Numerically, we discretise time and $\textbf{q}$-space choosing the grid such that a higher resolution does not further change the results.
Analytically, we use a Magnus expansion in $1/\omega$. The energy spectrum, Berry-curvature and Chern number are computed from $\hat{H}_\mathrm{eff}$ using the usual methods employed for static Hamiltonians. 
By comparing with numerical results and computing higher-order terms we show that the Magnus expansion can be truncated at first order in $1/\omega$ for our parameters, and $\hat{H}_\mathrm{eff}$ then takes on the form of the Haldane Hamiltonian (Eq. \ref{eq:HaldaneHam}).

There, $\Delta_{\mathrm{AB}}$ is not affected by the modulation compared to the static lattice and $t_{ij}$ are renormalized by a factor of about 0.85 with variations of $\pm 0.03$, depending on $\varphi$ and the orientation of the tunnelling.
The induced imaginary next-nearest-neighbour tunnelling takes on values of up to $h \times  18$\,Hz for the diagonal links and $h \times 5$\,Hz for the vertical links. Its value depends only on the modulation amplitude (scaling as a product of two second-order Bessel functions), the frequency (scaling as $1/\omega$), and the product of two nearest-neighbour tunnel couplings which correspond to this next-nearest-neighbour tunnelling.
Static real next-nearest neighbour tunnel couplings do not affect other terms in $\hat{H}_\mathrm{eff}$ and are not required to open a gap, meaning that our approach works equally well in deep optical lattices.
The weights $w_l$ determining the gap in Eq. (\ref{eq:gap}) are 3.5 and 2.1 for the vertical and diagonal tunnel couplings respectively, with variations of about $\pm0.1$ as a function of $\varphi$. Detailed derivations, formulae for all terms in the effective Hamiltonian, and comparisons between numerical and analytical approaches can be found in the Supplementary Material.

\subsection{Semi-classical calculations}

We use a semi-classical approximation to simulate the orthogonal drifts observed for Bloch oscillations along the $y$-direction in the experiment, using the same lattice parameters as in Fig.2. We denote the energy and Berry-curvature of the lowest band of the analytical effective Hamiltonian with $\epsilon (q_x,q_y)$ and $\Omega(q_x,q_y)$. The external accelerating force is given by $F_y=\pm 2\Delta E/\lambda$, where $\Delta E/h=114.6$ Hz is the energy offset per site. The equations of motion then read (omitting the $z$-direction, which de-couples)
\begin{eqnarray}
\dot{x} &=& \frac{1}{\hbar} \partial_{q_x} \epsilon (q_x,q_y) - \dot{q_y} \cdot \Omega(q_x,q_y)  \\
\dot{y} &=& \frac{1}{\hbar} \partial_{q_y} \epsilon (q_x,q_y) + \dot{q_x} \cdot \Omega(q_x,q_y)   \\
\hbar \dot{q_x} &=& -\partial_x V_{\mathrm{trap}}(x,y,z) \\
\hbar \dot{q_y} &=& F_y - \partial_y V_{\mathrm{trap}}(x,y,z) .
\label{eq:sc}
\end{eqnarray}
The effect of the harmonic trap is taken into account with $V_{\mathrm{trap}}(x,y,z)=0.5m(\omega_x^2x^2+\omega_y^2y^2+\omega_z^2z^2)$. 

As the underlying band structure possesses several symmetries, the possible values of the differential drift $\mathcal{D}$ are strongly constrained by the topology of the lowest band. When only TRS (IS) is broken, the band structure is point symmetric, $\epsilon (q_x,q_y)= \epsilon (-q_x,-q_y)$. Additionally,  the Berry-curvature is point symmetric (point anti-symmetric), so the lowest band is topologically non-trivial (trivial). If the system is also reflection symmetric, $\epsilon (q_x,q_y)= \epsilon (-q_x,q_y)$, the equation of motion for $\dot{q_x}$ remains unchanged when inverting the direction of the force in the topologically trivial case. Thus, a differential drift can only appear if the lowest band is topologically non-trivial. In the experiment, these symmetries are strictly present when $\varphi=\pm90^\circ$. The modulation weakly breaks reflection symmetry otherwise (see Supplementary Material) but it can be restored by considering the average of $\mathcal{D}(\varphi)$ and $\mathcal{D}(\pi-\varphi)$. For both cases the experimental data of Figs, 2d and 4c shows that the lowest band is topologically non-trivial.

For the numerical simulations, we compute $4\times 10^4$ trajectories, starting from a zero-temperature fermionic phase-space distribution (taking into account the $z$-direction) resembling the measured initial momentum-space distribution, and then determine the  $\textbf{q}$-space center-of-mass position after one Bloch cycle. For a rough estimate of the effects of Landau-Zener transfers to the second band, we record the minimum band-gap experienced by each trajectory and exclude trajectories below a chosen cut-off value. This approach will not capture the complex quantum-mechanical dynamics of the real transfer process, but may serve to indicate in which direction the measured drift-curves will be deformed.

\begin{figure}%
\includegraphics[width=\columnwidth]{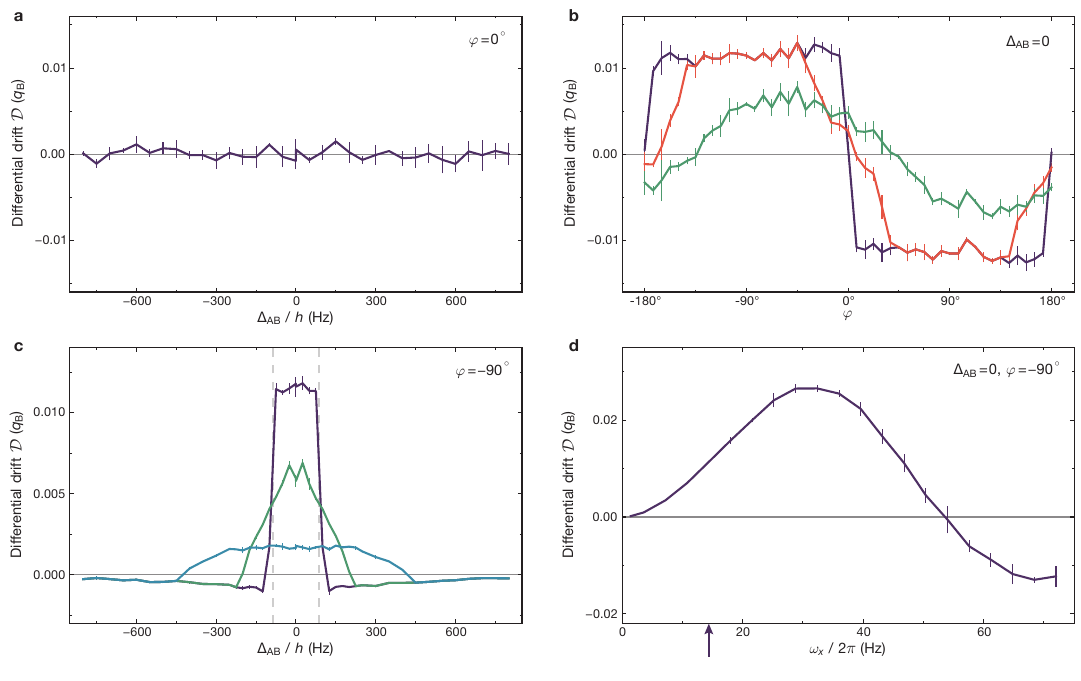}
\caption{{\bf Semi-classical simulations of the atomic motion.} The experiments shown in the main text in Fig. \ref{fig:observables}c and d are simulated using the semi-classical equations of motion (Eqs. 6-9). The band structure and the Berry-curvature are those of the effective Haldane Hamiltonian (Eq. 1). The atomic ensemble is modeled by a zero-temperature Fermi distribution. Data are mean $\pm$ s.d. of three simulations containing $4\times 10^4$ trajectories. The differential drift $\mathcal{D}$ is computed when breaking either IS ($\textbf{a}$) or TRS ($\textbf{b}$). The former shows no differential drift, in agreement with the experimental data of Fig. \ref{fig:observables}c. For the latter, the different curves take into account the transfer to the higher band by excluding trajectories passing through regions where the band-gap lies below a certain threshold. If this transfer is not taken into account (purple line), the differential drift varies sharply around $\varphi=0^{\circ}$ where the Chern number changes. However, as the threshold is raised to $0.5\,\Delta E $ (red line) and $\Delta E $ (green line) where $\Delta E /h = 114.6$\,Hz is the energy offset per site driving the Bloch oscillation, this sharp feature progressively smoothens and qualitatively reproduces the experimental measurements. $\textbf{c}$, When TRS is maximally broken ($\varphi=90^{\circ}$) and $\Delta_{\mathrm AB}$ varies, the transfer is also responsible for the differential drift extending beyond the topological phase. Without any transfer (purple line), the differential drift changes sharply around the topological phase transition (vertical dashed line), while it extends significantly in the topologically trivial phase when the threshold is set at $\Delta E$ (green line) or $3\,\Delta E$ (blue line). $\textbf{d}$, Influence of the transverse trapping frequency $\omega_x/2\pi$. The frequency used in the experiment is indicated by a purple arrow. For much larger frequencies the differential drift can vanish, as the transverse oscillation time becomes comparable to the Bloch period.}
\label{fig:edf1}%
\end{figure}

The results are shown in Extended Data Figure \ref{fig:edf1}. Panel a shows that no differential drift is expected when only IS is broken, as observed in the experiment. In particular, even though reflection symmetry is weakly broken in the system as stated above, its effect remains smaller than the numerical error on $\mathcal{D}$. The simulations furthermore show that for each individual applied force a drift should appear, as observed (see below and Extended Data Figure \ref{fig:edf2}a). When only TRS is broken, a differential drift which changes sign with the modulation phase $\varphi$ is computed, which is smaller but comparable to the measured values. The sudden change of $\cal{D}$ around $\varphi = 0^{\circ}$ is smoothed when taking into account transfer to higher bands. In that region the gap at the Dirac points and the spread of the Berry-curvature is very small, meaning that atoms which would contribute most to the drift are likely to be transferred to the higher band. A similar effect appears in Panel c, where both IS and TRS are broken for circular modulation. Without transfer, our simulations predict a sudden change in $\cal{D}$ at the topological transition. When taking transfer into account, the Dirac point with the smaller gap contributes less, so the drifts observed in the topological regime extend beyond the transition-line, as measured in the experiment (see Fig. \ref{fig:driftphasediagram}a). This transfer could be reduced by applying weaker gradients to the atomic cloud, which would however require removing the harmonic trapping potential along the $y$-direction. 

The underlying harmonic trap in the $x$-direction is also of particular importance, as it is responsible for transforming displacements in real space into momentum-space drifts. As seen in panel d, the differential drift vanishes when $\omega_x$ does, increases rapidly close to the experimental value of $\omega_x = 2\pi \times 14.4(6)$\,Hz and shows oscillatory behaviour when the time-scale of dipole oscillations becomes comparable to the Bloch-oscillation period.

\subsection{Drift measurements for opposite forces}

In this section, we discuss the individual drift observed after one full Bloch cycle supplementing Fig. \ref{fig:observables}c/d and Fig. \ref{fig:driftphasediagram} in the main text.
Our measurement technique probes the Berry-curvature of the lowest band by moving atoms in $q_{y}$-direction past the gapped Dirac points, where the Berry-curvature is localized, and relies on the coupling of real and momentum space.
Note that the width of the fermionic cloud is sufficiently large. By performing a Bloch oscillation we therefore sample the entire Brillouin-zone.  
As described in the main text, the applied uniform force causes an orthogonal velocity in real space when atoms are in the region of the Berry-curvature. 
The resulting change in real space position induces a transverse force arising from the underlying harmonic confinement in opposite direction of the displacement. 
As a result, we observe a drift in quasi-momentum, which is amplified by the negative effective mass at each Dirac point arising from the negative curvature of the band structure. 

\begin{figure}
\centering
\includegraphics[width=\columnwidth]{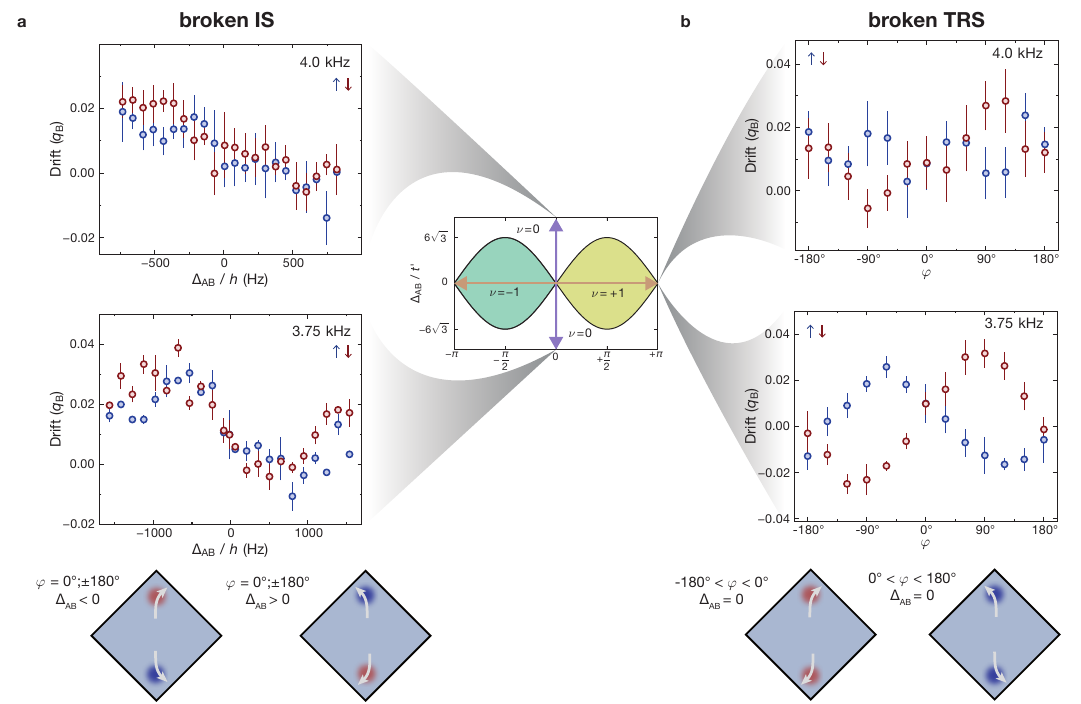}%
\caption{
{\bf Drift measurement for broken IS and TRS.} 
The measured drift used to obtain the differential Drift $\mathcal{D}$ in Fig. \ref{fig:observables}c and Fig. \ref{fig:driftphasediagram} of the main text is  individually shown for positive and negative forces in the $q_{y}$-direction. 
Data for positive (negative) force is shown in blue (red).
The central plot, showing the Haldane phase diagram, indicates the region which is scanned when breaking either IS (purple arrow) or TRS (brown arrow) in our system.  
{\bf a,} We break IS by introducing a sublattice offset and show measurements with modulation frequency of 4.0 kHz and 3.75 kHz.
Although the opposite Berry-curvatures at the two Dirac points sum up to zero within the first Brillouin-zone (BZ), we clearly see a drift depending on the size of $\Delta_{\mathrm{AB}}$.
Data show mean $\pm$ s.d. of at least 6 (4.0 kHz) or 2 (3.75 kHz) measurements. 
{\bf b,} By changing the modulation phase difference $\varphi$ we break TRS and the system enters the topologically non-trivial regime.
Drift data for positive (negative) force is shown in blue (red) for a modulation frequency of 4.0 kHz and 3.75 kHz. 
Data show mean $\pm$ s.d. of at least 21 (4.0 kHz) or 6 (3.75 kHz) measurements.
Schematics below show the expected orthogonal drifts caused by driving the atoms through the Berry curvature distribution. Red (blue) indicates positive (negative) Berry curvature. 
If only IS is broken {\bf(a)} the Berry curvature distribution is point-antisymmetric and changes sign when changing the sign of the sublattice offset.  
For opposite forces this leads to the same direction of the drift, as indicated by the white arrows.
If only TRS is broken {\bf(b)} the Berry curvature distribution at each Dirac point has the same sign, which is changed when reverting the rotation direction. 
In this case the opposite forces lead to opposite directions of the drift.   
}
\label{fig:edf2}
\end{figure}

In the topologically trivial case, when only IS is broken, for each individual gradient we observe an equal drift along $q_{x}$ when scanning the sublattice offset $\Delta_{\mathrm{AB}}$ (see Extended Data Figure \ref{fig:edf2}a).
This constitutes a measure of the local Berry curvature as the integrated Berry curvature is zero. 
We measure a drift which increases with increasing gap and changes sign with $\Delta_{\mathrm{AB}}$.
The data shows that the observed drift predominately arises from the first Dirac point which is passed. 
This effect can be explained in the following way:
The Berry curvature of the first Dirac point already leads to an orthogonal drift in quasi-momentum. 
When successively reaching the second Dirac point the shifted part of the cloud then does not experience the same Berry curvature distribution.
As expected, opposite oscillation directions give rise to the same drift, since not only the direction of the force changes but also the sign of the Berry curvature corresponding to the first Dirac point on the trajectory (see Extended Data Figure \ref{fig:edf3}).
Subtracting the drifts of opposite gradients gives the differential drift $\mathcal{D}$, as shown in the main manuscript. 
This probes the net contribution of both Dirac points.
Only for the largest $\Delta_{\mathrm{AB}}$ the measured drift decreases again, indicating an increasing spread of the Berry-curvatures distributions at each Dirac point, which then start to overlap and gradually cancel each other. 
We also observe these drifts in a static lattice with $\Delta_{\mathrm{AB}} \neq 0$.

In contrast, when only TRS is broken, we probe the topologically non-trivial regime. Extended Data Figure \ref{fig:edf2}b shows opposite drifts along $q_{x}$ for each of the oscillation directions.
In this case both successively passed Dirac points cause a drift in the same direction since the Berry curvature is point-symmetric. 
Therefore, changing the sign of the applied gradient leads to a drift in the opposite $q_{x}$-direction. 
As expected, the drift changes sign for the opposite modulation phase difference $\varphi$, directly revealing the changing sign of the Berry curvature distribution.
Here too, a larger gap (when $\varphi$ is closer to $\pm 90^{\circ}$ or the modulation frequency is lower) leads to a larger drift. 
The combination of the dependence of the drift on the size of the gap and the predominance of the first Dirac point on the trajectory may be the cause for the precursors seen in Fig. \ref{fig:driftphasediagram}a for simultaneously broken IS and TRS.

\subsection{Heating of a repulsively interacting Fermi gas}

\begin{figure}%
\centering
\includegraphics[width=\columnwidth]{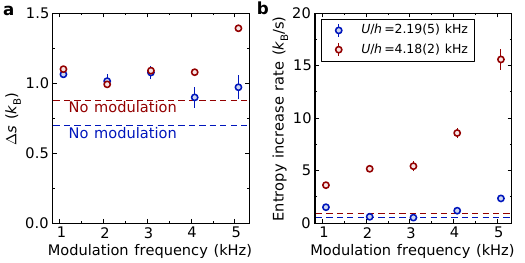}%
\caption{{\bf Heating of a repulsively interacting Fermi gas.}
{\bf a}, Entropy increase associated with loading into the modulated lattice and reverting the loading procedure. {\bf b}, Entropy increase rate in the modulated lattice for long holding times. The modulation frequency $\omega=2\pi\times 1080$\,Hz opens a gap of $h \times 44$\,Hz in the non-interacting band-structure. This value, in units of the tunnelling, is similar to the measurements of the main text. The dashed lines show the measured heating in a lattice without modulation with identical interaction strengths.}%
\label{fig:edf3}%
\end{figure}

We have also investigated the increase in entropy in the modulated honeycomb lattice in comparison to the static case by loading a repulsively interacting Fermi gas into the lattice and reverting the loading procedure \cite{Greif2011}. We prepare about $2.0(2)\times10^5$ atoms in a balanced spin mixture of the $|F,m_F\rangle=|9/2,-9/2\rangle$ and $|9/2,-5/2\rangle$ Zeeman states. The atoms are then loaded into a lattice with final depths $V_{\overline{\mathrm{X}},\mathrm{X},\mathrm{Y},\tilde{\mathrm{Z}}}=[14.0(4),0.79(2),\allowbreak 6.45(20),7.0(2)]\,E_{\mathrm{R}}$ within $200$\,ms. This corresponds to a system consisting of coupled isotropic honeycomb layers with a nearest-neighbour tunnelling $t/h=172(20)$\,Hz, as used in previous work \cite{Uehlinger2013}. The two lowest bands have a total bandwidth of $h \times 1.0$\,kHz, with a gap of $h \times 14$\,kHz to the next higher band (excluding the third direction). After turning on the lattice modulation (using the same $K_0$ as in the main text, which opens a gap of about $h \times 44$\,Hz at the Dirac points), we reverse the loading procedure and measure the final temperature of the sample. This is compared to the case where the lattice is not modulated. From the difference in temperature before loading the lattice and after the procedure, the corresponding entropy increase can be determined. We measure the entropy increase for different interaction strengths and modulation frequencies. In the Mott-insulating regime with $U/h=4.18(2)$\,kHz ($U/5t=4.9(6)$) and for a frequency of $\omega/2\pi=1.08$\,kHz (which is the same in proportion to the bandwidth as in the measurements of the main text) we find an entropy increase that is $25\%$ larger when modulating the lattice compared to the situation without modulation, see Extended Data Fig. \ref{fig:edf3}a. In the crossover regime at $U/h=2.19(5)$\,kHz ($U/5t=2.5(3)$) we find the same final entropy. This now correpsonds to a $40\%$ increase, which possibly originates from the creation of low-energy charge excitations for these parameters. In the measurements versus frequency, we cover the doublon excitation peak in the insulating phase, whose frequency is given by $U/h=4.18(2)$ kHz. We have furthermore measured the additional heating induced by holding the atoms in the modulated lattice for longer times, see Extended Data Fig. \ref{fig:edf3}b. For all parameters we find a linear increase of entropy with time. For timescales relevant for studying dynamics as in the measurements of the main text, this contribution is much smaller than the one associated with the modulation ramp. We find that the heating induced by the modulation does not dominate the final temperature, thus demonstrating that the scheme is well suited for studying many-body states in topological lattices.

\subsection{Proposal for creating spin-dependent Hamiltonians}

To realise spin-dependent Hamiltonians, e.g. the Kane-Mele model, we propose to use an oscillating magnetic gradient in order to apply an oscillating spin-dependent force on the atoms. For the $|9/2,-9/2\rangle$ state of $^{40}$K, maximum gradients of about 10\,G/cm would be required to achieve the same $K_0$ as in the main text when modulating with a frequency of 1\,kHz. Owing to the Zeeman effect, another spin component or atomic species would experience a different force. If the magnetic moment of the second component is chosen to be opposite, the Kane-Mele model is simply realised by replacing one oscillating mirror in the experiment by this oscillating gradient. The two spin components would then experience clockwise or anti-clockwise modulated forces, respectively, and therefore the two spin bands would have opposite Chern numbers. In general, a combination of an oscillating mirror and magnetic gradient can be used to create the desired average and differential force for other combinations of magnetic moments. This approach can be extended to bosonic atoms or Bose-Fermi mixtures and is also suitable for creating other types of spin-dependent tunnelling, e.g. situations where one species is pinned to the lattice and the other remains itinerant.

\section*{Supplementary material}
\setcounter{section}{0}
\setcounter{subsection}{0}
\setcounter{figure}{0}
\setcounter{equation}{0}
\setcounter{table}{0}
\renewcommand{\thefigure}{S\arabic{figure}} 
\renewcommand{\theequation}{S\arabic{equation}} 
\renewcommand\thetable{S\arabic{table}}
\renewcommand{\figurename}[1]{FIG. }

In the following we outline the theoretical framework used to obtain effective Hamiltonians for time-modulated optical lattices. In particular, we derive the mapping from an elliptically modulated honeycomb lattice to the Haldane Hamiltonian \cite{Haldane1988}. We consider a numerical and analytical approach, compare the results for a wide range of parameters and examine the validity of several approximations for the system studied in the experiment. Some elements of the general framework used there can be found in references \cite{Eckhardt2005, Oka2009, Kolovsky2011, Lebrat2013, Delplace2013, Goldman2014, Bukov2014}, and applications to circularly modulated honeycomb lattices can be found in very recent work \cite{Lebrat2013, Grushin2014, Zheng2014}.

\subsection{Effective Hamiltonian of a time-periodic system}
The evolution, given by time-evolution operator $\hat{U}$,  of a state obeying a time-periodic Hamiltonian of period $T$
is well captured by an effective Hamiltonian $\hat{H}_\text{eff}$ over timescales
greater than $T$ \cite{Shirley1965, Sambe1973}. An effective Hamiltonian is defined as,
assuming here and henceforth that $\hbar = 1$:
\begin{equation} \label{eq:time_dependent_heff}
\hat{U}(\tau+T, \tau) = \exp \left(-i \hat{H}_\text{eff}^\tau T \right) .
\end{equation}
The operator $\hat{H}_\text{eff}^\tau$ is known as the Floquet Hamiltonian.
By construction, its energy spectrum does not depend
on the choice of starting time $\tau$ as two time-evolution operators
with different starting times $\tau$, $\tau'$ are related
through a similarity transformation:
\begin{align} 
\hat{U}(\tau'+T, \tau') &= \hat{U}(\tau'+T, \tau+T) \hat{U}(\tau+T, \tau) \hat{U}(\tau, \tau') \nonumber \\
    &= \hat{U}(\tau', \tau) \hat{U}(\tau+T, \tau) \hat{U}(\tau', \tau)^{-1}
\end{align}
and so are two different effective Hamiltonians,
\begin{equation}
\hat{H}_\text{eff}^{\tau'} = \hat{U}(\tau', \tau) \hat{H}_\text{eff}^{\tau} \hat{U}(\tau', \tau)^{-1} . \label{eq:tautransformation}
\end{equation}

The effective Hamiltonian is proportional to the logarithm
of the time-evolution operator over a period $T$;
therefore its spectrum, known as the quasi-energy spectrum, is only defined modulo $\omega = 2\pi/T$.
This logarithm can be evaluated numerically,
as detailed in a later section (Eq. (\ref{eq:evol_steps}) and following).
It can alternatively be expanded as a Magnus series
involving multiple integrals and commutators of the time-dependent Hamiltonian:
up to first order,
\begin{equation}
\hat{H}_{\text{eff}}^\tau = \hat{H}_{0\omega} + \hat{H}_{1\omega}^\tau + \mathcal{O}\left(\frac{1}{\omega^2}\right)
\end{equation}
with
\begin{eqnarray}
\hat{H}_{0\omega} &=& \frac{1}{T} \int_\tau^{\tau+T} dt \hat{H}(t) \\
\hat{H}_{1\omega}^\tau &=& -\frac{i}{2 T} \int_\tau^{\tau+T} dt \int_\tau^t dt' [\hat{H}(t), \hat{H}(t')].
\end{eqnarray}
\, \\
Writing $\hat{H}(t)$ as a Fourier series,
\begin{eqnarray}
\hat{H}(t) = \sum_{n=-\infty}^{+\infty} \hat{H}_n e^{i n \omega t}
\end{eqnarray}
we compute:
\begin{eqnarray}
\hat{H}_{0\omega} &=& \hat{H}_0\\
\hat{H}_{1\omega}^\tau &=& \frac{1}{\omega} \sum_{n = 1}^{\infty} \frac{1}{n} \Big([\hat{H}_n, \hat{H}_{-n}] \label{eq:order1withT} \\
&&- e^{i n \omega \tau} [\hat{H}_n, \hat{H}_0] + e^{-i n \omega \tau} [\hat{H}_{-n}, \hat{H}_0] \Big) .\nonumber \\
\nonumber 
\end{eqnarray}

To lowest order, the effective Hamiltonian equals the average of the Hamiltonian over one period,
while the starting time $\tau$ only enters at first order in $1/\omega$
as a phase factor $e^{\pm i n \omega \tau}$. The information about the starting phase of the modulation may not be relevant
in a number of experimental cases -- for example when 
adiabatically switching on the modulation, as in the experiment we report on.
In a different approach \cite{Rahav2003, Goldman2014, Bukov2014},
the $\tau$-dependence of $\hat{H}_\text{eff}^\tau$ can
be absorbed by choosing a particular interaction picture,
splitting the evolution into an effective evolution under the Hamiltonian $\hat{H}_\text{eff}$
and two initial and final ``kicks'' defined by a $T$-periodic operator $\hat{K}(\tau)$
which averages to zero over one period:
\begin{equation}
\hat{U}(T+\tau, \tau) = e^{i K(\tau)} e^{-i \hat{H}_\text{eff} T} e^{-i K(\tau)} .
\end{equation}
The $\tau$-independent effective Hamiltonian can then be expanded up to $1/\omega^2$ as follows:
\begin{widetext}
\begin{align}
\hat{H}_{\text{eff}} &= \hat{H}_{0\omega} + \hat{H}_{1\omega} + \hat{H}_{2\omega} + \mathcal{O}\left(\frac{1}{\omega^3}\right) \\
\text{with} \quad \hat{H}_{0\omega} &= \hat{H}_0 \label{eq:order0noT} \\
\hat{H}_{1\omega} &= \frac{1}{\omega} \sum_{n = 1}^{\infty} \frac{1}{n} [\hat{H}_n, \hat{H}_{-n}] \label{eq:order1noT} \\
\hat{H}_{2\omega} &= \frac{1}{2 \omega^2} \sum_{n = 1}^{\infty} \frac{1}{n^2} \Big( [[\hat{H}_n, \hat{H}_0], \hat{H}_{-n}] + \text{h.c.} \Big)  \label{eq:order2noT} \\
& + \frac{1}{3 \omega^2} \sum_{n, n' = 1}^{\infty} \frac{1}{n n'} \Big( [\hat{H}_n, [\hat{H}_{n'}, \hat{H}_{-n-n'}]] - 2 [\hat{H}_n, [\hat{H}_{-n'}, \hat{H}_{n'-n}]] + \text{h.c.} \Big) .\nonumber 
\end{align}
\end{widetext}

The results presented above are general and valid for any time-periodic Hamiltonian. 

\subsection{Elliptical modulation}

We now consider the case of an optical lattice described by a tight-binding model
with one orbital per site.
The modulation applied in our experiments consists in moving the lattice
along a periodic trajectory $\vec{r}_\text{lat}(t)$,
giving rise to an inertial force $\vec{F}(t) = -m \ddot{\vec{r}}_\text{lat}(t)$
exerted on the atoms.
In the lattice frame, this amounts to adding a linear, site-dependent potential
to the tight-binding Hamiltonian at rest:
\begin{equation}
\hat{H}_\text{lat}(t) = \sum_{\langle i j \rangle} t_{ij} \hat{c}^\dagger_{i} \hat{c}_{j} + \sum_i \big(\vec{F}(t) \cdot \vec{r}_i \big) \, \hat{c}^\dagger_{i} \hat{c}_{i}
\end{equation}
where $\hat{c}_{i}$, $\hat{c}^\dagger_i$ denote the annihilation and creation operators
on the lattice sites at positions $\vec{r}_i$ in the lattice frame,
and $t_{ij}$ the tunnelling amplitudes of the static lattice.
This additional time-dependent term can in turn be cancelled by shifting the quantum states
by the lattice momentum $- \vec{q}_\text{lat} = -m \dot{\vec{r}}_\text{lat}(t)$, through the unitary transformation
\begin{equation} \label{eq:unitary_transformation}
\quad \hat{U}(t) = \exp \Big[i\sum_i \big(-m \dot{\vec{r}}_\text{lat}(t) \cdot \vec{r}_i \big) \, \hat{c}^\dagger_{i} \hat{c}_{i}\Big].
\end{equation}
In the resulting interaction Hamiltonian, this momentum shift is absorbed
in complex phase factors of the tunnelling amplitudes,
\begin{equation} \label{eq:int_hamiltonian}
\hat{H}'_\text{lat} = \hat{U}^\dagger \hat{H}_\text{lat} \hat{U} - i \hat{U}^\dagger \partial_t \hat{U}
= \sum_{\langle i j \rangle} e^{i \vec{q}_\text{lat}\cdot\vec{r}_{ij}} t_{ij} \hat{c}^\dagger_{i} \hat{c}_{j}
\end{equation}
where we define the relative vector between two sites $\vec{r}_{ij} = \vec{r}_i - \vec{r}_j$
and we omit writing the time-dependence of the operators and the lattice momentum for brevity. 
$\hat{H}_\text{eff}$ and associated dressed operators
and density matrices are the same in the laboratory frame as in the moving frame \cite{Bukov2014}.
The elliptical trajectories considered in this paper are created by modulating
the lattice position along two orthogonal axes $(\vec{e}_1, \vec{e}_2)$
with equal amplitude $A$ and a relative phase $\varphi$,
\begin{equation}
\vec{r}_{\text{lat}}(t) = -A \Big( \cos(\omega t) \vec{e}_1 + \cos(\omega t - \varphi) \vec{e}_2 \Big).
\end{equation}
Changing $\varphi$ allows for continuously changing both the aspect ratio of the trajectory
(from linear, $\varphi = 0^\circ$ or $180^\circ$, to circular, $\varphi = \pm 90^\circ$)
and its rotation direction (anticlockwise for $0^\circ < \varphi < 180^\circ$, clockwise for $-180^\circ < \varphi < 0^\circ$).
The phase factors in Eq. (\ref{eq:int_hamiltonian}), which essentially indicate
how the lattice velocity $\dot{\vec{r}}_\text{lat}$ projects onto the lattice bonds $\vec{r}_{ij}$,
are therefore sinusoidal functions of time: introducing the modulation parameters
\begin{align}
&\rho_{ij} e^{i \phi_{ij}} = \vec{r}_{ij}\cdot\vec{e}_1 + \vec{r}_{ij}\cdot\vec{e}_2 ~ e^{- i \varphi} \nonumber \\
&z_{ij} = m \omega A\,\rho_{ij} \label{eq:def_mod_par}
\end{align}
with the convention $\rho_{ij} \geq 0$, the Hamiltonian takes on the general form
\begin{equation} \label{eq:ham_phase_sin}
\hat{H} = \sum_{\langle i j \rangle} e^{i z_{ij} \sin(\omega t + \phi_{ij})} t_{ij} \hat{c}^\dagger_i \, \hat{c}_j.
\end{equation}
Its Fourier-harmonics, as required in Eqs. (\ref{eq:order0noT}) to (\ref{eq:order2noT}),
are obtained using the Jacobi-Anger expansion:
\begin{equation} \label{eq:ham_bessel}
\hat{H}_n = \sum_{\langle i j \rangle} J_n(z_{ij}) e^{i n \phi_{ij}} t_{ij} \hat{c}^\dagger_i \, \hat{c}_j
\end{equation}
where $J_n$ denotes the $n^\text{th}$-order Bessel function of the first kind.
Every $\hat{H}_n$ features the same space periodicity and underlying geometry as the tight-binding lattice at rest,
with modified tunnelling amplitudes and phases.
The Bessel functions of negative order are related to the positive ones through $J_{-n}(z) = (-1)^n J_n(z)$.
In the limit of weak modulations $(z_{ij} \ll 1)$, we can write 
$J_n(z_{ij}) = z_{ij}^n/(2^n n!) + \mathcal{O}(z_{ij}^{n+2})$
and limiting the expansion of $\hat{H}_\text{eff}$ to the lowest orders in $n$ is justified.
This no longer holds for stronger modulations, where care has to be taken how the  truncation is performed.
Note that so far no assumptions about the lattice geometry have been made.

\subsection{Tight-binding model of a bipartite lattice}
\begin{figure}
\includegraphics[width=1.09\columnwidth]{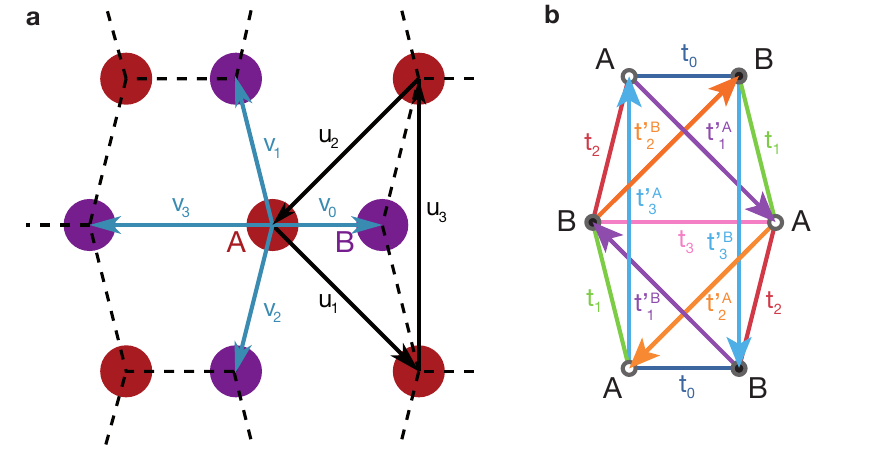}
\caption{
\textbf{a}, Definition of the Bravais lattice vectors $\vec{u}_1$, $\vec{u}_2$, intra-sublattice vector $\vec{u}_3$ and the inter-sublattice vectors 
$\vec{v}_0, \vec{v}_1, \vec{v}_2, \vec{v}_3$
for the honeycomb lattices considered in the experiment. \textbf{b}, Tunnelling structure. Arrows indicate the definition of phases $\Phi_{j'}$.
}
\label{fig:bravais_lattice}
\end{figure}
The honeycomb lattice used in the experiment contains two sites per unit cell,
belonging to two different chequerboard sublattices $\mathcal{A}$ and $\mathcal{B}$,
see Fig. \ref{fig:bravais_lattice}a. We now allow for an energy difference $\Delta_\text{AB}$ between the two sublattices, which is not affected by the unitary transformation of Eq. (\ref{eq:unitary_transformation}), 
and distinguish the tunnellings $t_j$ connecting $\mathcal{A}$ and $\mathcal{B}$ on the one hand,
and the tunnellings $t'^A_{j'}$ and $t'^B_{j'}$ within $\mathcal{A}$ and $\mathcal{B}$ on the other hand.
The resulting tight-binding Hamiltonian reads:
\begin{align}
\hat{H} = \sum_{\vec{u} \in \mathcal{A}} \Big[&
      \frac{\Delta_\text{AB}}{2} \, \hat{a}_{\vec{u}}^{\dagger} \hat{a}_{\vec{u}}
    - \frac{\Delta_\text{AB}}{2} \, \hat{b}_{\vec{u}+\vec{v}_0}^{\dagger} \hat{b}_{\vec{u}+\vec{v}_0} \\
    + & \sum_j (t_j \hat{b}_{\vec{u}+\vec{v}_j}^{\dagger} \hat{a}_{\vec{u}} + \text{h.c.}) \nonumber \\
    + & \sum_{j'} (t'^A_{j'} \hat{a}_{\vec{u}+\vec{u}_{j'}}^{\dagger} \hat{a}_{\vec{u}}
    + t'^B_{j'} \hat{b}_{\vec{u}+\vec{v}_0-\vec{u}_{j'}}^{\dagger} \hat{b}_{\vec{u}+\vec{v}_0} + \text{h.c.}) \Big] \nonumber
\end{align}
where the vectors $\vec{v}_j$ connect $\mathcal{A}$\,--$\mathcal{B}$ site pairs
and the vectors $\vec{u}_{j'}$ (which include the Bravais lattice vectors) connect $\mathcal{A}$\,--$\mathcal{A} / \mathcal{B}$\,--$\mathcal{B}$ site pairs.
Here, $\hat{a}_{\vec{r}}$, $\hat{a}^\dagger_{\vec{r}}$ ($\hat{b}_{\vec{r}}$, $\hat{b}^\dagger_{\vec{r}}$) denote the annihilation and creation operators on a site belonging to the $\mathcal{A}$ ($\mathcal{B}$) sublattice.
The Hamiltonian has been chosen such that the complex $t'^A_{j'}$ and $t'^B_{j'}$ have their phases defined along the arrows shown in Fig. \ref{fig:bravais_lattice}b.
Taking the Fourier tranform of the annihilation and creation operators on the sublattices,
\[
\hat{a}_\vec{q} = \frac{1}{\sqrt{N}} \sum_{\vec{u} \in \mathcal{A}} e^{-i \vec{q}\cdot\vec{u}} \hat{a}_\vec{u}, \quad
\hat{b}_\vec{q} = \frac{1}{\sqrt{N}} \sum_{\vec{u}' \in \mathcal{B}} e^{-i \vec{q}\cdot\vec{u}'} \hat{b}_{\vec{u}'}
\]
the tight-binding Hamiltonian can be rewritten in quasi-momentum space as:

\begin{align}
\hat{H}(\vec{q}) &=
(\hat{a}^\dagger_\vec{q} \,\, \hat{b}^\dagger_\vec{q})
\begin{pmatrix}
    h_{AA} & h_{AB}^* \\
    h_{AB} & h_{BB}
\end{pmatrix}
\begin{pmatrix}
    \hat{a}_\vec{q} \\
    \hat{b}_\vec{q}
\end{pmatrix}
\nonumber \\
&= h_i \hat{I} + h_x \hat{\sigma}_x + h_y \hat{\sigma}_y+ h_z \hat{\sigma}_z
\label{eq:Pauli_coeff_tb}
\end{align}

~ \\

where we define the operators $\hat{O} = (\hat{a}^\dagger_\vec{q}, \hat{b}^\dagger_\vec{q}) \, O \, (\hat{a}_\vec{q}, \hat{b}_\vec{q})^T$
acting on the space spanned by the Bloch waves residing on the two sublattices
$\hat{a}^\dagger_\vec{q} \ket{0}$ and $\hat{b}^\dagger_\vec{q} \ket{0}$,
with $I$ the $2 \times 2$ identity matrix and $\sigma_{x,y,z}$ the Pauli matrices satisfying the commutation relations
$[\sigma_\alpha, \sigma_\beta] = 2i \epsilon_{\alpha \beta \gamma} \sigma_\gamma$.
The coefficients $h_{i,x,y,z}$ are expressed as:
\begin{widetext}
\begin{align} 
h_i &= \sum_{j'} \Re (t'^A_{j'} + t'^B_{j'}) \cos(\vec{q}\cdot\vec{u}_{j'}) + \Im (t'^A_{j'} - t'^B_{j'}) \sin(\vec{q}\cdot\vec{u}_{j'}) \\
h_x &= \Re\big(\sum_j t_j e^{i\vec{q}\cdot\vec{v}_j}\big), \quad h_y = \Im\big(\sum_j t_j e^{i\vec{q}\cdot\vec{v}_j}\big) \\
h_z &= \frac{\Delta_\text{AB}}{2} +  \sum_{j'} \Re (t'^A_{j'} - t'^B_{j'}) \cos(\vec{q}\cdot\vec{u}_{j'}) + \Im (t'^A_{j'} + t'^B_{j'}) \sin(\vec{q}\cdot\vec{u}_{j'}) \label{eq:Pauli_coeff_tb_sigmaz}
\end{align}
\end{widetext}
The energies of the associated energy bands are 
\begin{equation}
\epsilon_\pm(\vec{q}) = h_i \pm \sqrt{h_x^2 + h_y^2 + h_z^2} .
\end{equation}
~\\

The Dirac points are located at the quasi-momenta $\vec{q}_D$ where
$h_x(\vec{q}_D) = h_y(\vec{q}_D) = 0$, which is a necessary condition
for band degeneracy, $\epsilon_+(\vec{q}_D) = \epsilon_-(\vec{q}_D)$.
If all tunnel couplings $t_j$ are real, every Dirac point $\vec{q}_D$ is paired with
another one at opposite quasi-momentum $-\vec{q}_D$,
as $h_x(-\vec{q}_D) = h_x(\vec{q}_D)$ and $h_y(-\vec{q}_D) = -h_y(\vec{q}_D)$.
In the case where $t'^A_{j'} = t'^B_{j'} = t'_{j'} e^{i \Phi_{j'}}$ for all $j'$,
the gaps at two opposite Dirac points $\pm \vec{q}_D$ are given by
\begin{equation}
G_\pm = \epsilon_+(\pm\vec{q}_D) - \epsilon_-(\pm\vec{q}_D) = |\Delta_\text{AB} \pm \Delta_\text{T}|.
\end{equation}
$|\Delta_\text{T}|$ is the gap induced by the complex tunnellings
when inversion symmetry is preserved ($\Delta_\text{AB} = 0$),
\begin{equation} \label{eq:delta_t}
\Delta_\text{T} = - \sum_{j'} w_{j'} t'_{j'} \sin (\Phi_{j'}).
\end{equation}
It is the sum of the imaginary parts of the complex amplitudes $t'_{j'} e^{i \Phi_{j'}}$
weighted by $w_{j'} = -4 \sin(\vec{q}_D.\vec{u}_{j'})$.
The weights are positive for the lattice used in our experiment, but can also be negative.
Their norms are sensitive to the position of the Dirac points, given by the identity $\sum_j t_j e^{\vec{q}_D.\vec{v}_j} = 0$, and therefore depend on the
inter-sublattice tunnel couplings $t_j$ on the one hand,
and the vectors $\vec{v}_j$ setting the geometry of the lattice
on the other hand.

In the experiment, the amplitude of $t_3$, which corresponds to a next-next-nearest-neighbour tunnelling, can in general be significant. However, it does not qualitatively change the band-structure of the system. Its main contribution is to move the position of the Dirac points, in the same way as a larger value of $t_0$ would.

\subsection{Analytical effective Hamiltonian}
We now derive an analytical expression of the effective Hamiltonian of the lattice under elliptical modulation in the approach exposed earlier.
We assume that the next-nearest couplings of the static lattice are real and show $\mathcal{A} - \mathcal{B}$ symmetry, $t'^A_{j'} = t'^B_{j'} = t'_{j'}$. This symmetry may be broken experimentally but the size of this effect is negligible for the lattices considered in this work.
The harmonics of the modulated Hamiltonian, whose tunnel couplings are modified according to Eq. (\ref{eq:ham_bessel}),
can be transformed in momentum space $\vec{q}$ as $\hat{H}_n = h_{ni} \hat{I} + h_{nx} \hat{\sigma}_x + h_{ny} \hat{\sigma}_y+ h_{nz} \hat{\sigma}_z$,
with
\begin{widetext}
\begin{align}
h_{ni} &= \sum_{j'} 2 J_n(z_{j'}) t_{j'} \cos(\vec{q}\cdot\vec{u}_{j'}) \\
h_{nx} &= \sum_{j} \frac{1}{2} [e^{i \vec{q}\cdot\vec{v}_j} + (-1)^n e^{-i \vec{q}\cdot\vec{v}_j}] J_n(z_j) e^{i n \phi_j} t_j \\
h_{ny} &= \sum_{j} \frac{1}{2i} [e^{i \vec{q}\cdot\vec{v}_j} - (-1)^n e^{-i \vec{q}\cdot\vec{v}_j}] J_n(z_j) e^{i n \phi_j} t_j \\
h_{nz} &= \frac{\Delta_\text{AB}}{2} \delta_n.
\end{align}
\end{widetext}
Note that $h_{nx}$ and $h_{ny}$ are complex numbers, which accounts for the fact that $\hat{H}_n = \hat{H}_{-n}^\dagger$
is in general not Hermitian. $\delta_n$ denotes the Kronecker delta.
Inserting these expressions into Eqs. (\ref{eq:order0noT}) to (\ref{eq:order2noT}) results in
\begin{widetext}
\begin{align}
\hat{H}_{0\omega} &= \sum_{j'} 2 t'_{0,j'} \cos(\vec{q}\cdot\vec{u}_{j'}) \hat{I}
+ \sum_j t_{0,j} \big[ \cos(\vec{q}\cdot\vec{v}_j) \hat{\sigma}_x + \sin(\vec{q}\cdot\vec{v}_j) \hat{\sigma}_y \big]
+ \frac{\Delta_\text{AB}}{2} \hat{\sigma}_z \label{eq:eff_order0} \\
\hat{H}_{1\omega} &= \sum_{j_1 > j_2} 2 t_{1, j_1 j_2} \sin (\vec{q}\cdot\vec{v}_{j_1 j_2}) \, \hat{\sigma}_z \label{eq:eff_order1} \\
\hat{H}_{2\omega} &= \sum_{j_1, j_2, j_3} t_{2, j_1 j_2 j_3}  \big[
\cos \, \vec{q}\cdot(\vec{v}_{j_1 j_2} - \vec{v}_{j_3}) \, \hat{\sigma}_x + \sin \, \vec{q}\cdot(\vec{v}_{j_1 j_2} - \vec{v}_{j_3}) \, \hat{\sigma}_y \big] \nonumber\\
& + \sum_{j_1, j_2, j_3} t_{2, j_1 j_2 j_3}  \big[
\cos \, \vec{q}\cdot(-\vec{v}_{j_1 j_2} - \vec{v}_{j_3}) \, \hat{\sigma}_x
+ \sin \, \vec{q}\cdot(-\vec{v}_{j_1 j_2} - \vec{v}_{j_3}) \, \hat{\sigma}_y \big]
+ \sum_{j_1 > j_2} 2 t_{2, j_1 j_2}  \cos (\vec{q}\cdot\vec{v}_{j_1 j_2}) \, \hat{\sigma}_z
\end{align}
\begin{align}
t_{0,j} &= J_0(z_j) \, t_j \qquad t'_{0,j'} = J_0(z_{j'}) \, t'_{j'} \label{eq:t0} \\
t_{1, j_1 j_2} &= \frac{2 t_{j_1} t_{j_2}}{\omega} \sum_{n=1}^\infty \frac{(-1)^n}{n} \sin [n(\phi_{j_1} - \phi_{j_2})] J_n(z_{j_1}) J_n(z_{j_2}) \label{eq:t1}\\
t_{2, j_1 j_2} &= \frac{2 \Delta_\text{AB} t_{j_1} t_{j_2}}{\omega^2} \sum_{n=1}^\infty \frac{1}{n^2} \cos [n(\phi_{j_1} - \phi_{j_2})] J_n(z_{j_1}) J_n(z_{j_2}) \label{eq:t22}\\
t_{2, j_1 j_2 j_3} &= \frac{2t_{j_1} t_{j_2} t_{j_3}}{\omega^2} \Big[\sum_{n \text{ odd}} \frac{1}{n^2} \cos [n(\phi_{j_1} - \phi_{j_2})] J_n(z_{j_1}) J_n(z_{j_2}) J_0(z_{j_3})
- \sum_{n \text{ even}} \frac{i}{n^2} \sin [n(\phi_{j_1} - \phi_{j_2})] J_n(z_{j_1}) J_n(z_{j_2}) J_0(z_{j_3}) \Big] \label{eq:t23}
\end{align}
\end{widetext}
At the lowest order, $\hat{H}_{0 \omega}$,
all inter-sublattice tunnel couplings $t_{0,j}$
and intra-sublattice tunnel couplings $t_{0,j'}$
are renormalized according to the zeroth order Bessel function $J_0(z_j)$
while the sublattice offset $\Delta_\text{AB}$ remains unaffected. 

The first-order term $\hat{H}_{1 \omega}$ adds a purely imaginary part $it'_{1, j_1 j_2}$ (cf. Equation (\ref{eq:Pauli_coeff_tb_sigmaz}))
to the preexisting next-nearest-neighbour (NNN) couplings,
now indexed by pairs of sites $(j_1, j_2)$
connected by the vector $\vec{v}_{j_1 j_2} = \vec{v}_{j_1} - \vec{v}_{j_2}$.
Under this form, the new tunnellings may be interpreted as a succession of two virtual tunnel processes
with amplitudes $J_0(z_{j_1}) \, t_{j_1}$ and $J_0(z_{j_2}) \, t_{j_2}$.
Note that the size of this induced NNN tunnelling depends only on the size of the static NN tunnelling, meaning that the scheme does not require shallow lattices.

The second order term $\hat{H}_{2 \omega}$ leads to new complex NN and NNNN tunnellings $t_{2, j_1 j_2}$,
as well as a real $\mathcal{A} - \mathcal{B}$ imbalance $\pm t_{2, j_1 j_2 j_3}$ of the NNN tunnellings if $\Delta_\text{AB} \neq 0$.
The static next-nearest-neighbour couplings $t'_{j'}$ do not appear anywhere but in the zeroth order term $t_{0,j'}$,
and therefore do not affect the energy-splitting between the two bands.

In general, prefactors to the identity matrix in Eq. (\ref{eq:Pauli_coeff_tb})
will not contribute to higher-order terms in the effective Hamiltonian,
as the commutators $[I, \sigma_{\alpha}]$ in Eqs. (\ref{eq:order1withT}), (\ref{eq:order1noT}) and (\ref{eq:order2noT}) vanish.
Importantly, this means that the next-nearest-neighbour tunnellings of the static lattice
do not affect the position of the topological transition lines.
In absence of interactions, the presence of the higher-order terms crucially relies on
the presence of two bands or more for the commutators in Eqs. (\ref{eq:order1withT}), (\ref{eq:order1noT}) and (\ref{eq:order2noT}) not to be zero.
The higher-order terms hence vanish when considering non-interacting atoms on lattices with a single orbital per unit cell.

For completeness, we also provide the $\tau$-dependent
term at $1/\omega$ order, although it is not used, as explained above. The expression may however be relevant for experiments where a modulation is suddenly switched on.
Denoting the additional term containing the commutators $[\hat{H}_{\pm n}, \hat{H}_0]$
in Eq. (\ref{eq:order1withT}) as $\hat{H}_{1'\omega}^\tau$, we find
\begin{widetext}
\begin{align}
\hat{H}_{1'\omega}^\tau &=
\sum_j t_{1,j}^\tau \big[ \cos(\vec{q}\cdot\vec{v}_j) \hat{\sigma}_x + \sin(\vec{q}\cdot\vec{v}_j) \hat{\sigma}_y \big]
+ \bigg\{ \frac{\Delta^\tau_1}{2}
+ \sum_{j_1 > j_2} \Big[ 2 t^{\tau \, \Re}_{1, j_1 j_2} \cos (\vec{q}\cdot\vec{v}_{j_1 j_2})
+ 2 t^{\tau \, \Im}_{1, j_1 j_2} \sin (\vec{q}\cdot\vec{v}_{j_1 j_2}) \Big] \bigg\} \hat{\sigma}_z \\
t_{1,j}^\tau &= - \sum_{n \text{ odd}} \frac{2 \Delta_\text{AB} t_j}{n \omega} J_n(z_j) \cos [n (\phi_j + \omega \tau)]
- i \sum_{n \text{ even}} \frac{2 \Delta_\text{AB} t_j}{n \omega} J_n(z_j) \cos [n (\phi_j + \omega \tau)] \label{eq:nn_tau} \\
\Delta^\tau_1 &= \sum_{n \text{ odd}} \sum_j \frac{2 t_j^2}{n \omega} J_n(z_j) J_0(z_j) \cos [n (\phi_j + \omega \tau)] \label{eq:delta_tau} \\
t^{\tau \, \Re}_{1, j_1 j_2} &= \sum_{n \text{ odd}} \frac{2 t_{j_1} t_{j_2}}{n\omega} \Big[ J_n(z_{j_1}) J_0(z_{j_2}) \cos [n (\phi_{j_1} + \omega \tau)]
+ J_0(z_{j_1}) J_n(z_{j_2}) \cos [n (\phi_{j_2} + \omega \tau)] \Big] \label{eq:nnn_re_tau} \\
t^{\tau \, \Im}_{1, j_1 j_2} &= -\sum_{n \text{ even}} \frac{2 t_{j_1} t_{j_2}}{n\omega} \Big[ J_n(z_{j_1}) J_0(z_{j_2}) \sin [n (\phi_{j_1} + \omega \tau)]
- J_0(z_{j_1}) J_n(z_{j_2}) \sin [n (\phi_{j_2} + \omega \tau)] \Big] \label{eq:nnn_im_tau}
\end{align}
\end{widetext}
In addition to (\ref{eq:eff_order0}) and (\ref{eq:eff_order1}),
the effective $\tau$-dependent Hamiltonian features complex NN tunnel couplings $t_{1,j}^\tau$,
a sublattice offset $\Delta^\tau_1$ and complex NNN tunnel couplings
$t^{\tau \, \Re}_{1, j_1 j_2} + i t^{\tau \, \Im}_{1, j_1 j_2}$.

\subsection{Numerical effective Hamiltonian}
For all numerical calculations, $\hat{H}(t)$ is approximated by an operator $\hat{H}(t_i)$ that is
piece-wise constant on $N$ consecutive time intervals $[t_i, t_{i+1}[$, where $t_i = i \,  T / N$, $i = 0 \, ... \, N-1$.
This enables us to rewrite the time-evolution operator over $[0, T[$
as the product of $N$ shorter time-evolution operators:
\begin{equation} \label{eq:evol_steps}
    \hat{U}(T, 0) = \prod_{i=0}^{N-1} {U}(t_{i+1}, t_i) = \prod_{i=0}^{N-1} e^{-i\hat{H}(t_i)T/N} .
\end{equation}
$\hat{H}(t_i)$ is evaluated for every $\vec{q}$ separately according to Eq.\,(\ref{eq:ham_phase_sin}). 
The effective Hamiltonian $\hat{H}^0_\text{eff}$ is then computed from Eq. (\ref{eq:time_dependent_heff}),
\begin{equation}  \label{eq:numerical_basis}
\hat{H}^0_\text{eff} =\frac{i}{T} \log{\hat{U}(T, 0)} .
\end{equation}
Its eigenvectors $\ket{u^\pm_\vec{q}}$ and energies $\epsilon(\vec{q})$ are further
used to compute the Berry curvature for the lowest energy band\cite{Xiao2010}:
\begin{equation}
\Omega(\vec{q}) = 2 \Im \Bigg[ \frac{\braket{u_{\vec{q}}^- | \partial_{q_1} \hat{H}^0_\text{eff}(\vec{q}) | u_{\vec{q}}^+ }
\braket{u_{\vec{q}}^+ | \partial_{q_2} \hat{H}^0_\text{eff}(\vec{q}) | u_{\vec{q}}^- } }
{(\epsilon_+(\vec{q})-\epsilon_-(\vec{q}))^2} \Bigg]
\end{equation}
approximating the partial derivatives along the axes of the Brillouin zone
by their first order finite-difference expressions.
The Chern number is obtained by integrating the Berry curvature over the entire Brillouin zone:
\begin{equation}
\nu = \frac{1}{2\pi} \int_{\vec{q} \in \mathrm{BZ}} d^2 \vec{q} \, \Omega_-(\vec{q}),
\end{equation}
where the integral is replaced by a sum over a discrete grid.
The grid spacing and number of time steps are always chosen such
that a higher resolution does not change the results we obtain any further. 
The numerical calculations do include $\tau$-dependent terms,
and we have verified that changing $\tau$ does not affect the energy spectrum,
as expected from Eq. (\ref{eq:tautransformation}).

With this numerical approach the effect of other types of modulation can also be computed.
For example, we verify that a cyclic modulation of the tunnelling amplitudes
can also be used to create topological bands,
as shown in \cite{Kitagawa2010}.

\subsection{Results for an ideal brickwall lattice}

\begin{figure}[b]
\includegraphics[width=\columnwidth]{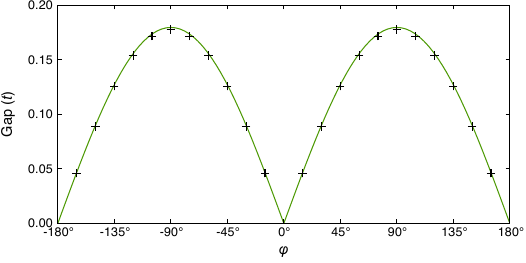} 
\caption{
Analytically and numerically computed gap versus relative phase $\varphi$ between horizontal and vertical modulation,
for an ideal brickwall lattice and $\omega = 10 t$. $K_0 = \pi^2\,(A/\lambda)\,(\hbar\,\omega/E_{\mathrm{R}}) = m \omega A \lambda/2 = 0.7778$ is the dimensionless parameter
for the modulation amplitude.
Here and in the following figures,
points (lines) indicate the numerical (analytical) results.
}
\label{fig:gap_vs_phi_ideal}
\end{figure}

We first consider an idealised version of the lattice realised in the experiment,
which exhibits nearly all features of the realistic lattice when applying elliptical modulation:
the balanced brickwall lattice, whose horizontal and vertical bonds are all of equal length
($|\vec{v_0}| = |\vec{v_1}| = |\vec{v_2}| = \lambda/2$), at square angle,
and which only includes the three main nearest-neighbour tunnel couplings,
with equal tunnelling amplitudes
($t_0 = t_1 = t_2 = t$ and $t_3 = t'_1 = t'_2 = t'_3 = 0$).
In the presence of inversion symmetry, $\Delta_\text{AB} = 0$,
an elliptical modulation opens an equal gap at both Dirac points,
located at quasi-momenta
$\pm \vec{q}_D = \mp 4 \pi / ( 3\lambda^2) \vec{u}_1 \mp 4 \pi / (3\lambda^2) \vec{u}_2$.
This fixes the weights present in Eq. (\ref{eq:delta_t})
to $w_1 = w_2 = w_3 = 2\sqrt{3}$.
Since $\vec{v}_0$ is horizontal and $\vec{v}_1, \vec{v}_2$ are vertical,
the modulation amplitudes $z_{ij}$ of the NN bonds (as defined in Eq. (\ref{eq:def_mod_par}))
are always $K_0 = \pi^2\,(A/\lambda)\,(\hbar\,\omega/E_{\mathrm{R}}) = m \omega A \lambda/2 = 0.7778$. Therefore the effective nearest-neighbour tunnellings are
\begin{equation}
t^{\mathrm{eff}}_0 = t^{\mathrm{eff}}_1 =t^{\mathrm{eff}}_2 = t J_0(K_0),\label{eq:ideal_teff}
\end{equation}
independent of $\varphi$.
The associated modulation phases are exactly $\phi_0 = 0$, $\phi_1 = -\varphi$ and $\phi_2 = \pi - \varphi$,
which allows for expressing the new complex NNN tunnelling (\ref{eq:t1}) as
\begin{align}
t'^{\mathrm{eff}}_1 &= i\frac{2 t^2}{\omega} \sum_{n=1}^\infty \frac{(-1)^n}{n} \sin (n \varphi) [J_n(K_0)]^2 \\
t'^{\mathrm{eff}}_2 &= - i\frac{2 t^2}{\omega} \sum_{n=1}^\infty \frac{1}{n} \sin (n \varphi) [J_n(K_0)]^2 \\
t'^{\mathrm{eff}}_3 &= 0.
\end{align}
Consequently, the energy offset caused by breaking TRS reads
\begin{align}
\Delta_\text{T} &= \frac{8 \sqrt{3} t^2}{\omega} \sum_{n\text{ odd}} \frac{1}{n} \sin (n \varphi) [J_n(K_0)]^2 \nonumber \\
&\approx \Delta_\text{T}^\text{max} \sin(\varphi). \label{eq:sin_phi}
\end{align}
The approximation corresponds to keeping only the \\ $n = 1$ term of the sum, with $\Delta_\text{T}^\text{max} = 8 \sqrt{3} [t J_1(K_0)]^2 / \omega$.

For $K_0 = 0.7778$ (as used in the experiment), this approximation deviates by less than 0.1\% from the asymptotic value of the maximum gap obtained for $n \rightarrow \infty$.
Fig. \ref{fig:gap_vs_phi_ideal} shows the gap as a function of $\varphi$ computed through
both numerical and analytic methods; they show overall excellent agreement
with a relative difference of approximately 1\%.

\begin{figure}
\includegraphics[width=\columnwidth]{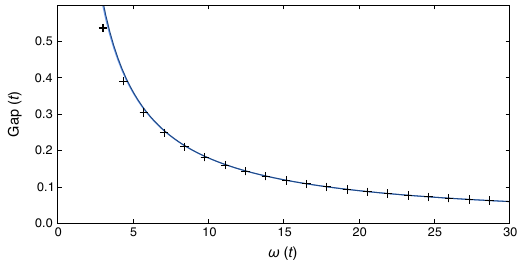}
\caption{
Absolute gap versus modulation frequency $\omega$, for circular modulation $\varphi = \pm 90^\circ$, $K_0 = 0.7778$. The dark (light) blue line shows the analytical results truncated at first (second) order in $1/\omega$. The two lines are almost identical.
}
\label{fig:gap_vs_omega_ideal}
\end{figure}

\begin{figure}
\includegraphics[width=\columnwidth]{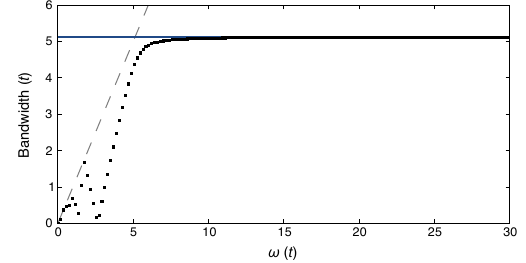}
\caption{
Bandwidth (defined as the energy splitting at $\vec{q} = 0$) versus modulation frequency $\omega$, for circular modulation $\varphi = \pm 90^\circ$,
$K_0 = 0.7778$. The dashed grey line indicates the identity line $f(\omega) = \omega$
which sets the maximum possible bandwidth of the quasi-energy spectrum.
}
\label{fig:bw_vs_omega_ideal}
\end{figure}

Figs. \ref{fig:gap_vs_omega_ideal} and \ref{fig:bw_vs_omega_ideal}
show the frequency dependence of the gap and the bandwidth of the effective
quasi-energy band-structure. Whereas the bandwidth remains close to its static value renormalized by
the zeroth-order Bessel function ($6 t J_0(K_0) \approx 5.13 t$ with $K_0 = 0.7778$), 
the gap increases as the modulation frequency $\omega$ decreases
and goes like $1 / \omega$ for $\omega \gg t$. We find only minor corrections
owing to the $1 / \omega^2$ term of the analytic expansion,
meaning that a truncation to $1^\text{st}$ order in $1 / \omega$ is acceptable for our parameters.
Deviations from the numerical calculations start to appear for frequencies below the static bandwidth $6t$.
The induced gap remains considerable even for the highest frequencies
that would realistically be employed in an experiment.
Truncating the effective Hamiltonian at $0^\text{th}$ order in $1/\omega$ is therefore
not a valid approximation for the honeycomb lattice.

\begin{figure}
\includegraphics[width=0.95\columnwidth]{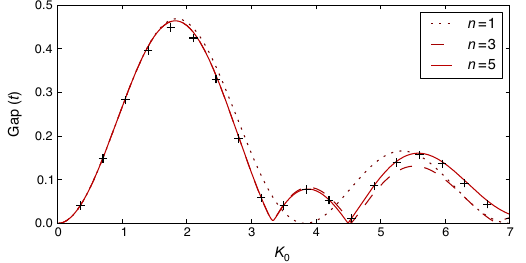}
\caption{
Absolute gap versus modulation amplitude $K_0$, for circular modulation $\varphi = \pm 90^\circ$,
$\omega = 10t$.
The analytic line is plotted at second order in $1/\omega$
and increasing number $n$ of the harmonics in the time-dependent Hamiltonian expansion. The gap changes sign around $K_0 = 4$
}
\label{fig:gap_vs_k0_ideal}
\end{figure}

\begin{figure}
\includegraphics[width=\columnwidth]{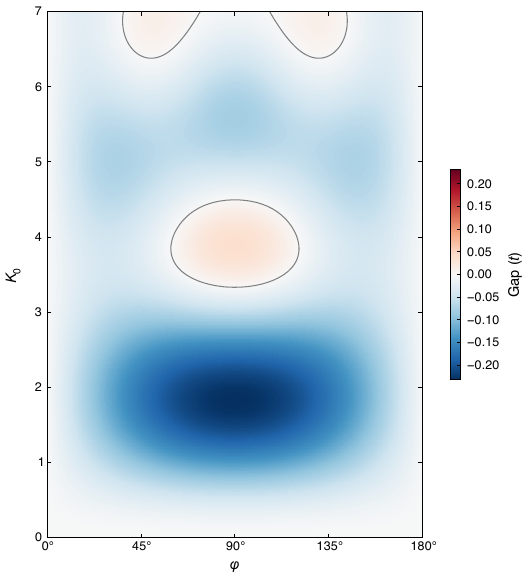}
\caption{%
Analytically computed gap versus relative phase $\varphi$ and modulation amplitude $K_0$
at $\omega = 10 t$.
Red indicates the $\nu = 1$ phase, blue the $\nu = -1$ phase. The gap closes on the solid grey line.
}
\label{fig:phase_diagram_honeycomb}
\end{figure}

The gap is a non-monotonic function of the modulation amplitude $K_0$.
It is well matched by the analytic expansion up to $n = 1$ for amplitudes below $K_0 \approx 2$, whilst for amplitudes up to $K_0 \approx 7$ expanding to at least $n = 5$ is necessary (Fig. \ref{fig:gap_vs_k0_ideal}).
Strikingly, the sign of the gap (and therefore the Chern number $\nu$) alternates for increasing $K_0$
(the first inversion occurs between $K_0 = 3.3$ and 4.5),
which is also \textit{not} predicted by the first order theory.
The general phase diagram in $(\varphi, K_0)$ space
is drawn in Fig. \ref{fig:phase_diagram_honeycomb},
and shows that, for certain large values of $K_0$, the $\nu = +1$ phase
becomes accessible for positive $\varphi$.

\begin{figure}
\includegraphics[width=\columnwidth]{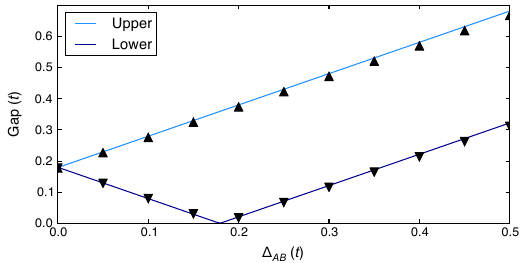}
\caption{
Absolute gap for the two Dirac points vs. sublattice offset $\Delta_\text{AB}$ with 
$\varphi = 90^\circ$, $\omega = 10t$, $K_0 = 0.7778$. The single vanishing gap signals the topological transition.
}
\label{fig:gap_vs_delta_ideal}
\end{figure}

With a preexisting sublattice offset $\Delta_\text{AB}$,
the gaps at the two Dirac points differ as soon as the modulation is switched on, $K_0 > 0$, as shown in Fig. \ref{fig:gap_vs_delta_ideal}.
One of them closes as soon as $\Delta_\text{AB} = \Delta_\text{T}$, as expected from the Haldane model (see also Eq. (3) of the main text).

\begin{figure}
\includegraphics[width=\columnwidth]{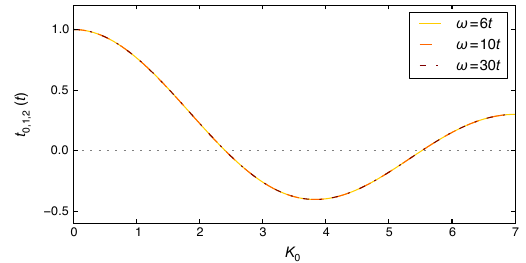}
\includegraphics[width=\columnwidth]{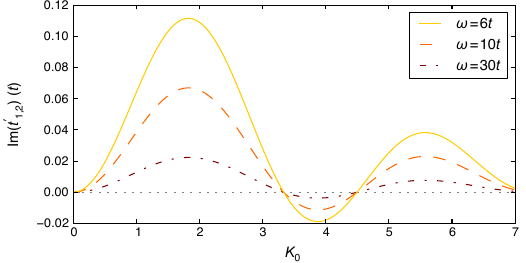}
\caption{
Tunnelling versus modulation amplitude $K_0$, for circular modulation $\varphi = 90^\circ$
at different frequencies $\omega = 30t, 10t, 6t$.
The contribution from the $1/\omega^2$ term to the tunnellings between $\mathcal{A}$ and $\mathcal{B}$ sublattices
is below $10^{-3} t$ at $\omega = 6t$.
}
\label{fig:tun_vs_k0_ideal}
\end{figure}

\begin{figure}
\includegraphics[width=\columnwidth]{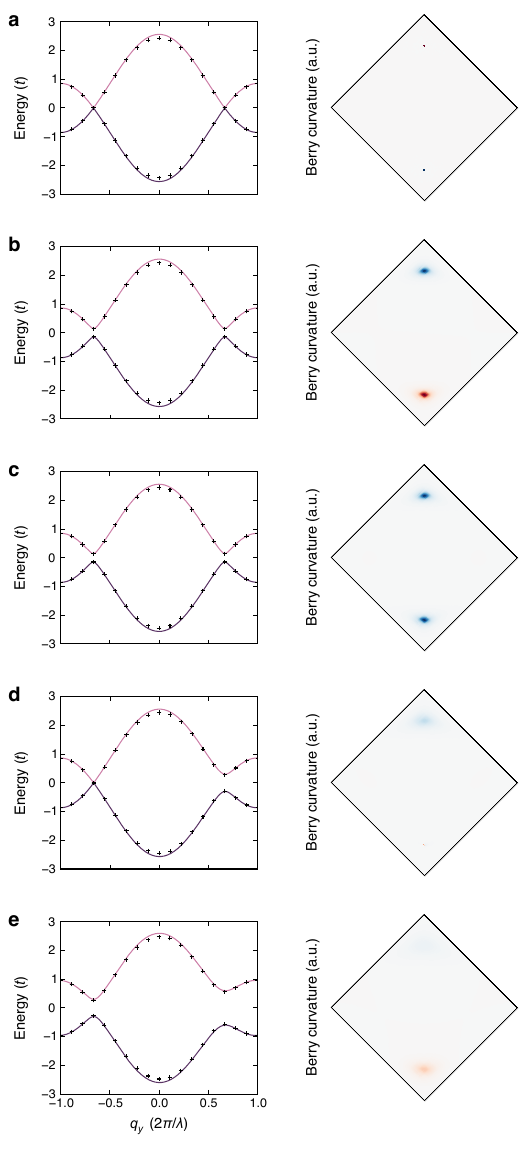}
\caption{%
Cut of the energy bands along the vertical line $q_x = 0$
and Berry curvature of the modulated ideal brickwall lattice
for $\omega = 6t$, $K_0 = 0.778$ and \\
\textbf{a.} $\varphi = 0$, $\Delta_\text{AB} = 0$ (IS and TRS not broken); \\
\textbf{b.} $\varphi = 0$, $\Delta_\text{AB} = \vert \Delta_\text{T}^{\mathrm{max}}\vert \approx 0.30t$ (trivial phase, $\nu = 0$); \\
\textbf{c.} $\varphi = 90^\circ$, $\Delta_\text{AB} = 0$ (non-trivial phase, $\nu = -1$); \\
\textbf{d.} $\varphi = 90^\circ$, $\Delta_\text{AB} = \Delta_\text{T}^{\mathrm{max}}$ (topological transition); \\
\textbf{e.} $\varphi = 90^\circ$, $\Delta_\text{AB} = 3\Delta_\text{T}^{\mathrm{max}}$ (trivial phase, $\nu = 0$).
Note that in \textbf{e}, the gap induced by IS is larger than the gap related to broken TRS, but the latter clearly influences the band-structure and Berry-curvature.
}
\label{fig:band_structures_ideal}
\end{figure}

While the effective energy bands do not depend on any particular choice of the modulation starting time $\tau$,
the tunnel couplings do when using the numerical approach (Eqs. (\ref{eq:nn_tau}) to (\ref{eq:nnn_im_tau})).
Fig. \ref{fig:tun_vs_k0_ideal} reports on the tunnellings as computed analytically
through Eqs. (\ref{eq:t0}) to (\ref{eq:t23}),
where the dependence is explicitly absent.
As stated in Eq. (\ref{eq:ideal_teff}) all NN tunnellings are evenly renormalized according to the zero-order Bessel function (Fig. \ref{fig:tun_vs_k0_ideal}), independent of $\omega$.
To a good approximation, the $t'_1, t'_2$ tunnellings are purely imaginary
and display the $\omega$, $K_0$ dependence of the gap highlighted in Figs. \ref{fig:gap_vs_omega_ideal} and \ref{fig:gap_vs_k0_ideal}. 
In Figure \ref{fig:band_structures_ideal} numerical and analytical results for the entire band-structure are shown for exemplary parameters,
as well as analytically computed Berry-curvature distributions.
Additionally, we confirm that the numerically computed Berry-curvature results in Chern numbers of $0.0$ and $\pm 1.0$ where expected.
Overall we see that the analytical solution, truncated at $1^{\mathrm{st}}$ order in $1/\omega$ provides reliable results for all experimentally relevant regimes.

\subsection{Results for the experimental honeycomb lattice}

\begin{table}
\begin{tabular}{c|c|c|c|c}
& Static & $\varphi=0^\circ$ & $\varphi=180^\circ$ & $\varphi=\pm 90^\circ$ \\
\hline
$t_0$ & $-746$ & $-662$ & $-662$ & $-662$ \\
\hline
$t_1$ & $-527$ & $-467$ & $-431$ & $-449$ \\
\hline
$t_2$ & $-527$ & $-431$ & $-467$ & $-449$ \\
\hline
$t_3$ & $-126$ & $-103$ & $-103$ & $-103$ \\
\hline
$t'_1$ & 14 & 14 & 7 & $10 \mp 18i$ \\
\hline
$t'_2$ & 14 & 7 & 14 & $10 \mp 18i$ \\
\hline
$t'_3$ & 61 & 29 & 29 & $29 \mp 5i$ \\

\end{tabular}
\caption{
Parameters of the static lattice used in the experiment (as frequencies $t/h$ in units of Hz),
obtained through an \emph{ab initio} computation of the Wannier functions \cite{Uehlinger2013},
along with their modified values in presence of linear or circular modulation
($\omega = 2\pi \times 4000$ Hz, $K_0 = 0.7778$), calculated analytically.
When going from one kind of linear modulation to the orthogonal one
($\varphi = 0^\circ \leftrightarrow 180^\circ$),
tunnelling energies are swapped according to a $x \mapsto -x$ reflection.
Reverting the circular modulation from clockwise ($\varphi=-90^\circ$)
to anticlockwise ($\varphi=+90^\circ$) replaces the next-nearest-neighbour couplings
by their complex conjugates. Furthermore, $|\vec{v}_0| = 0.438\,\lambda$.
\label{tab:parameters}
}
\end{table}

Compared to the idealised case treated above, in the lattice realised in the experiment the NN tunnelling are not all equal (and $t_3 \neq 0$),  meaning that the induced tunnel couplings and the weights contributing to the gap are also different.
Furthermore, the static real NNN tunnel couplings are not zero, which affects the shape of the band-structure, but not the energy difference between the two bands or the induced tunnelling, as outlined above.
Finally, the lattice has a slightly shorter lattice spacing $|\vec{v}_0| = 0.438\,\lambda$ along $x$, implying that the bond angle departs from $90^\circ$.
The tunnelling parameters of the static lattice used in this paper are reproduced in Tab. \ref{tab:parameters}.

\begin{figure}
\includegraphics[width=\columnwidth]{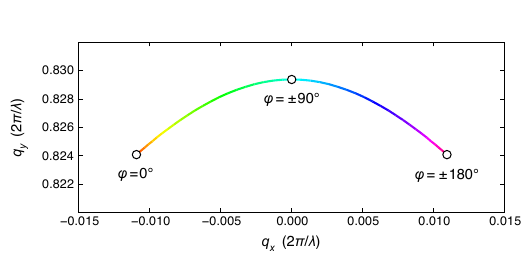} 
\caption{
Trajectory of one of the two Dirac points in quasi-momentum space
when $\varphi$ is varied, for $\omega = 2 \pi \times 4000$ Hz, $K_0 = 0.7778$
(the other Dirac point is located at opposite quasi-momentum).
}
\label{fig:kgap_vs_phi_real}
\end{figure}

\begin{figure}
\includegraphics[width=\columnwidth]{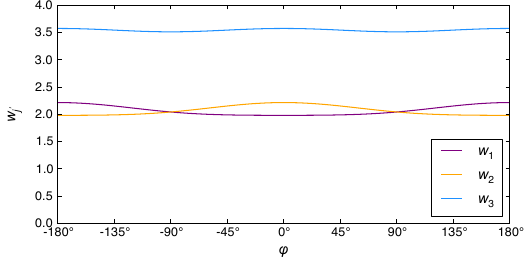} 
\caption{
Weights $w_{j'}$ relating imaginary next-nearest-neighbour tunnelling to the gap $\Delta_\text{T}$, as a function of $\varphi$.
}
\label{fig:w_vs_phi_real}
\end{figure}

\begin{figure}
\includegraphics[width=\columnwidth]{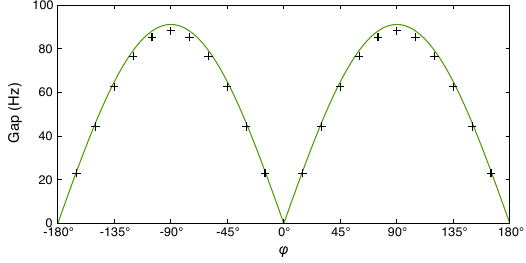} 
\caption{
Analytic (lines) and numerical (crosses) absolute gap of the experimental lattice
versus relative phase $\varphi$ between horizontal and vertical modulation,
for $\omega = 2 \pi \times 4000$ Hz, $K_0 = 0.7778$.
}
\label{fig:gap_vs_phi_real}
\end{figure}

\begin{figure}
\includegraphics[width=\columnwidth]{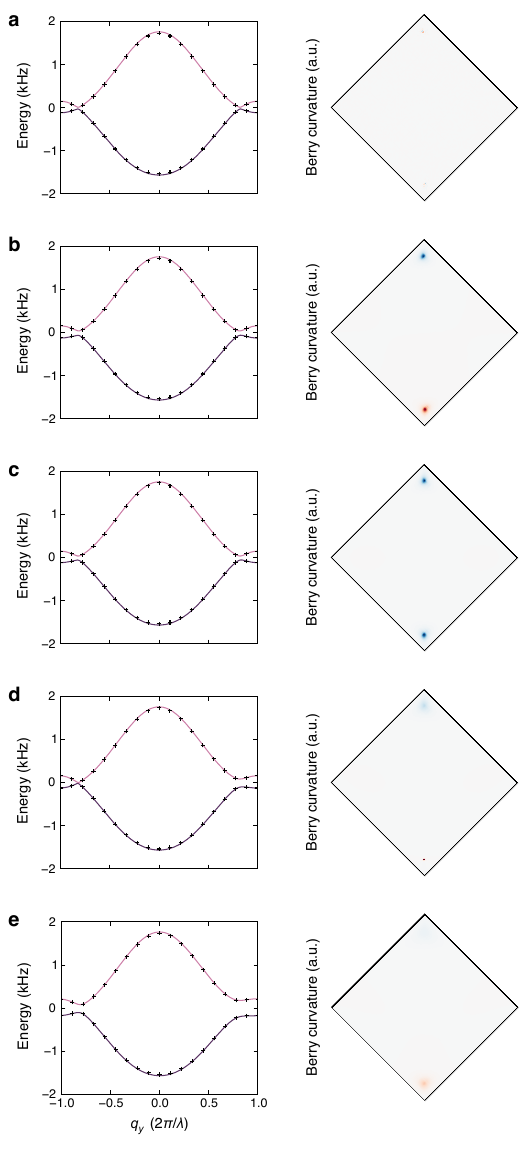}
\caption{%
Cut of the energy bands along the vertical line $q_x = 0$
and Berry curvature of the modulated realistic honeycomb lattice
for $\omega = 2 \pi \times 4000$ Hz, $K_0 = 0.7778$ and \\
\textbf{a.} $\varphi = 0$, $\Delta_\text{AB} = 0$ (IS and TRS not broken); \\
\textbf{b.} $\varphi = 0$, $\Delta_\text{AB} = \vert\Delta_\text{T}^{\mathrm{max}}\vert = h \times 88$ Hz (trivial phase, $\nu = 0$); \\
\textbf{c.} $\varphi = 90^\circ$, $\Delta_\text{AB} = 0$ (non-trivial phase, $\nu = -1$); \\
\textbf{d.} $\varphi = 90^\circ$, $\Delta_\text{AB} = \vert\Delta_\text{T}^{\mathrm{max}}\vert$ (topological transition); \\
\textbf{e.} $\varphi = 90^\circ$, $\Delta_\text{AB} = 3\vert\Delta_\text{T}^{\mathrm{max}}\vert$ (trivial phase, $\nu = 0$). \\
The Dirac points are rotated out of the vertical in panels \textbf{a}. and \textbf{b}.,
which translates to an apparent gap in the band cut of \textbf{a}.
and a slightly asymmetric Berry curvature in \textbf{b}.
}
\label{fig:band_structures_real}
\end{figure}

As a result of $|\vec{v}_0| \neq |\vec{v}_3|$, $t'^{\mathrm{eff}}_3 \neq 0$. Moreover, the NN tunnellings $t_j$ are renormalized slightly differently when modulating, as the projections in Eq. (\ref{eq:def_mod_par}) depend on the orientation and length of the tunnelling vectors. When the modulation trajectory is not circular, this weakly breaks the $x \mapsto -x$ reflection symmetry of the lattice, leading to a displacement of the Dirac points away from the $q_x = 0$ line
of the Brillouin-zone, see Fig. \ref{fig:kgap_vs_phi_real}. However for $\varphi = 0^{\circ}$ or $\pm 180^{\circ}$, this only amounts to a movement of the Dirac point by about  1\% of the Brillouin-zone size.
This effect creates a slight $\varphi$ dependence for the weights $w_{j'}$
in Eq. (\ref{eq:delta_t}), see Fig. \ref{fig:w_vs_phi_real}.

At a modulation frequency $\omega = 2 \pi \times 4000$ Hz and an amplitude $K_0 = 0.7778$,
the maximum gap is achieved for circular modulation and is numerically computed to be $h \times 88$\,Hz (analytically we find 91\,Hz). Deviations from the sinusoidal dependence on $\varphi$ owing to higher-order terms of the perturbative expansion remain below 2\%, see Fig. \ref{fig:gap_vs_phi_real}). The calculated band-structures and Berry-curvatures for our experimental parameters are shown in Fig. \ref{fig:band_structures_real}.
%
%

\end{document}